%% file: main.tex
\documentclass[twocolumn]{aastex63}

\usepackage{natbib}
\usepackage{CJK}

\usepackage{soul}

\newcommand\N[1]{{\color{orange} \bf #1}} 

\shorttitle{SN~2019esa}
\shortauthors{Andrews et al.}

\begin{document}

\title{High Cadence TESS and ground-based data of SN~2019esa, the less energetic sibling of SN~2006gy\footnote{This paper includes data gathered with the 6.5~m Magellan Telescope at Las Campanas Observatory, Chile.}}

\input{affiliation}
\input{authors}

\begin{abstract}

We present photometric and spectroscopic observations of the nearby ($D\approx28$~Mpc) interacting supernova (SN) 2019esa, discovered within hours of explosion and serendipitously observed by the \emph{Transiting Exoplanet Survey Satellite} (\emph{TESS}).  Early, high cadence light curves from both \emph{TESS} and the DLT40 survey tightly constrain the time of explosion, and show a 30 day rise to maximum light followed by a near constant linear decline in luminosity.
Optical spectroscopy over the first 40 days revealed a highly reddened object with narrow Balmer emission lines seen in Type IIn supernovae.
The slow rise to maximum in the optical lightcurve combined with the lack of broad H$\alpha$ emission  suggest the presence of very optically thick and close circumstellar material (CSM) that quickly decelerated the supernova ejecta. This CSM was likely created from a massive star progenitor with an $\dot{M}$ $\sim$ 0.3 M$_{\sun}$ yr$^{-1}$ lost in a previous eruptive episode 3--4 years before eruption, similar to giant  eruptions of luminous blue variable stars. At late times, strong intermediate-width \ion{Ca}{2}, \ion{Fe}{1}, and \ion{Fe}{2} lines are seen in the optical spectra, identical to those seen in the superluminous interacting SN~2006gy. The strong CSM interaction masks the underlying explosion mechanism in SN~2019esa, but the combination of the luminosity, strength of the H$\alpha$ lines, and mass loss rate of the progenitor all point to a core collapse origin.

\end{abstract}


\section{Introduction}
\label{sec:intro}
Core collapse supernovae (CCSNe) represent the end of the evolution of massive stars ($\gtrsim$ 8 M$_{\sun}$). Those SNe that exhibit hydrogen emission lines in their spectra are typically classified as Type II SNe, with further sub-type classifications depending on either their lightcurve shapes (Type II-L, Type II-P) or spectroscopic signatures (Type IIn, Type IIb). See \citet{2017hsn..book..239A,2017hsn..book..195G,2017suex.book.....B,2017hsn..book..403S} for detailed reviews. The class of CCSNe that show narrow or intermediate-width hydrogen emission lines in their spectra produced from the interaction between the SN ejecta and the surrounding circumstellar material (CSM) are classified as Type IIn \citep[where the ``n'' stands for narrow]{1990MNRAS.244..269S}. In these systems, the narrow lines ($\sim$10$^{2}$ km s$^{-1}$) are formed from the ionization of the slow moving CSM, while  intermediate-width lines of a few 10$^{3}$ km s$^{-1}$ arise from the interaction of the fast SN shock and the CSM. In many instances broad 10$^{4}$ km s$^{-1}$ emission is also seen, tracing the freely expanding SN ejecta.  The ejecta-CSM interaction can create interesting multi-component emission lines that change as the SN evolves \citep[e.g.][]{2000ApJ...536..239L,2015MNRAS.449.1876S,2017MNRAS.471.4047A}.

Type IIn SNe represent a diverse class of objects, and may  come from various progenitor systems.  Absolute magnitudes can range as high as --22 mag, as was seen in the superluminous IIn SN~2006gy \citep{2007ApJ...659L..13O,2007ApJ...666.1116S}, although they generally reach a maximum brightness of --17 to --19 mag \citep{2013A&A...555A..10T}.  This makes SN IIn brighter than most other CCSNe, owing to the kinetic energy of the SN explosion being converted to thermal energy through shock interaction. The progenitors of these SNe~IIn are likely special cases of evolved massive stars with pre-supernova outbursts, and could include extreme red supergiants (RSGs), yellow hypergiants (YHGs), or luminous blue variables (LBVs) \citep{2014ARA&A..52..487S}.

The timescale of interaction can be fleeting or long-lasting depending on the radial extent and density of the CSM, which is a direct product of the mass-loss history of the progenitor.  Estimates for CSM masses of most SNe IIn can range anywhere from 0.1 to 20 M$_{\sun}$ with mass-loss rates of  10$^{-4}$ to 1 M$_{\sun}$ yr$^{-1}$ \citep{2007ApJ...666.1116S,2012ApJ...744...10K}.   The CSM interaction not only affects the spectroscopic evolution but the lightcurve evolution as well, with rise-times to maximum ranging from 14 to 61 days \citep{2020A&A...637A..73N}. Some objects even have flat late-time lightcurves which maintain brightness for decades after eruption \citep[e.g. SN~1988Z and SN~2005ip]{1993MNRAS.262..128T,2020MNRAS.498..517F,2017MNRAS.466.3021S}. For a full review on interacting SNe see \citet{2017hsn..book..403S}.

Because a determining property of SN IIn is nearby CSM creating narrow emission lines, thermonuclear SNe Ia  can also create IIn observational signatures if they are surrounded by dense H-rich shells. Often referred to as Ia-CSM (or IIn/Ia), these objects show suggestive evidence of underlying SN Ia spectra, such as broad Fe and Si absorption lines diluted by excess continuum luminosity, as well as narrow hydrogen emission, and are often brighter than their normal SN Ia counterparts \citep{2006ApJ...650..510A,2003Natur.424..651H,2013ApJS..207....3S}.   While many SNe IIn have been posited to actually be SNe Ia-CSM, only a handful show unambiguous evidence for SN Ia features  due to the strength of the CSM interaction generally hiding the underlying Ia spectra. The best clear-cut case is that of PTF11kx which showed late-onset interaction \citep{2012Sci...337..942D,2013ApJ...772..125S}, but the list of potential SN Ia-CSM is growing \citep{2016MNRAS.459.2721I,2015MNRAS.447..772F,2019ApJ...871...62G}. However, the claim that these events actually arise from a thermonuclear SN Ia explosion has been controversial in several cases, since SNe Ic can have similar spectral appearance at some phases, especially when diluted by CSM interaction \citep{2006ApJ...653L.129B}.

\begin{figure}
    \centering
    \includegraphics[width=\linewidth]{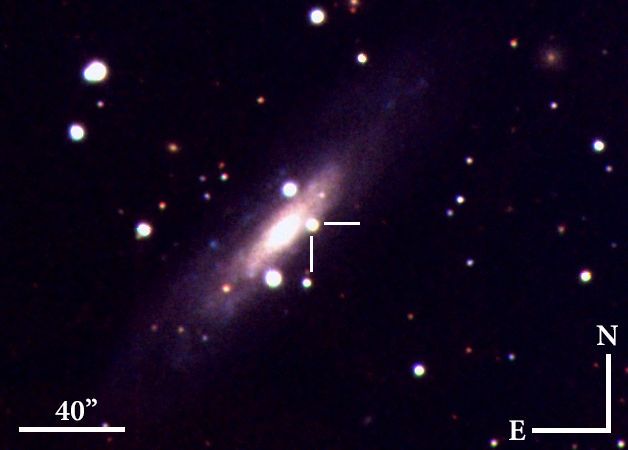}
    \caption{Composite $gri$ Las Cumbres Observatory Siding Springs image of SN~2019esa taken on 2019 May 8. }
    \label{fig:finder}
\end{figure}

Often,  non-terminal events such as luminous red novae (LRNe) or LBV outbursts are classified as Type IIn events due to the narrow H$\alpha$ component \citep{2021MNRAS.506.4715R}. Referred to as ``SN imposters" \citep{2000PASP..112.1532V,2011MNRAS.415..773S,2012ApJ...758..142K,VanDyk+2012,2019A&A...630A..75P}, their fainter absolute magnitudes at maximum light, $-10 < M_{V} < -15$ mag,  generally differentiate them from  terminal IIn events. In some cases the line between true supernova and SN imposter is blurry. For instance the terminal nature of both SN~2009ip and SN~2015bh are still debated \citep{2013MNRAS.433.1312F,2017A&A...599A.129T,2016MNRAS.463.3894E,2016MNRAS.463.2904S,2022arXiv220502896S}.

The optically thick CSM interaction zone in these objects masks the underlying ejecta, making
it difficult to determine the type of explosion, whether it be core collapse, thermonuclear, or even a non-terminal eruption.  One such object which falls into this category is the nearby SN~2019esa (Figure \ref{fig:finder}), which was discovered in ESO 035- G 018 at an RA(2000) $=$ 07$^h$55$^m$00$^s$.95, Dec(2000) $=-76\degr 24\arcmin 43\farcs 06$ during the course of the DLT40 one-day cadence SN search \citep[for a description of the survey, see][]{Tartaglia18} on 2019 May 06 \citep[MJD~58609.154]{Sand19esa}. This object was given the designation DLT19c by the DLT40 team, but we use the IAU naming convention and refer to it as SN~2019esa throughout this work. The discovery magnitude was $r$=17.2 mag, or $M_r \approx -$15.0 mag, given the distance modulus we adopt below, which is at the transition from SN imposter to real SN.  Within a day of discovery it was classified as a IIn due to the prominent narrow Balmer emission lines \citep{2019ATel12736....1U,2019TNSCR.738....1H}, and over the next month it rose to a maximum $r$=14.5 mag, or $M_r \approx -$17.7 mag, moving it into the regime of SNe IIn and not imposters.  SN~2019esa was in the \emph{Transiting Exoplanet Survey Satellite} (\emph{TESS}) footprint, constraining the explosion date to MJD~58608.44 \citep[2019 May 5.94]{2021MNRAS.500.5639V}, a value we adopt throughout the paper.

The most recent Tully-Fisher distance measurement of ESO~035--G018 is listed as 27.80 $\pm$ 1.49 Mpc, using a redshift of z=0.00589, and an  $H_{0}$ = 75.0 km s$^{-1}$ Mpc$^{-1}$ \citep{2016AJ....152...50T}.  This gives a $\mu$ = 32.22 $\pm$ 0.15 mag which we will use for the following analysis. We do note the large range of redshift-independent distances to this galaxy (21.0 Mpc $<$ D $<$ 32 Mpc) listed in NASA/IPAC Extragalactic Database (NED), which add uncertainties to the distance calibrated measurements. This paper is structured as follows: in \S\ref{sec:obs} observations and data reduction are outlined,  we discuss the photometric and bolometric evolution in \S\ref{sec:LC}, \S\ref{sec:specev} details the spectroscopic evolution of the object, in \S\ref{sec:Disc} we lay out the implications of the observational data, and finally the results are summarized in \S\ref{sec:Conc}.

\section{Observations and Data Reduction} \label{sec:obs}
\subsection{Imaging}

 Continuous photometric monitoring of SN~2019esa was done by the DLT40 survey's two discovery telescopes, the PROMPT5 0.4-m telescope at Cerro Tololo International Observatory and the PROMPT-MO 0.4-m telescope at Meckering Observatory in Australia, operated by the Skynet telescope network \citep{Reichart05}.  The PROMPT5 telescope has no filter (`Open') while the PROMPT-MO telescope has a broadband `Clear' filter, both of which we calibrate to the Sloan Digital Sky Survey $r$ band \citep[see ][for further reduction details]{Tartaglia18}.  The sky location of SN~2019esa is far enough in the South that it is circumpolar, and by keeping loose hour angle constraints we were able to continuously observe the SN even when it was difficult with other facilities. The last non-detection from DLT40 was two days before discovery, on 2019 May 04 (JD 2458607.65), or 31 hours before the estimated explosion, and the first DLT40 detection was roughly 17 hours after explosion.

\begin{figure*}
\includegraphics[width=\linewidth]{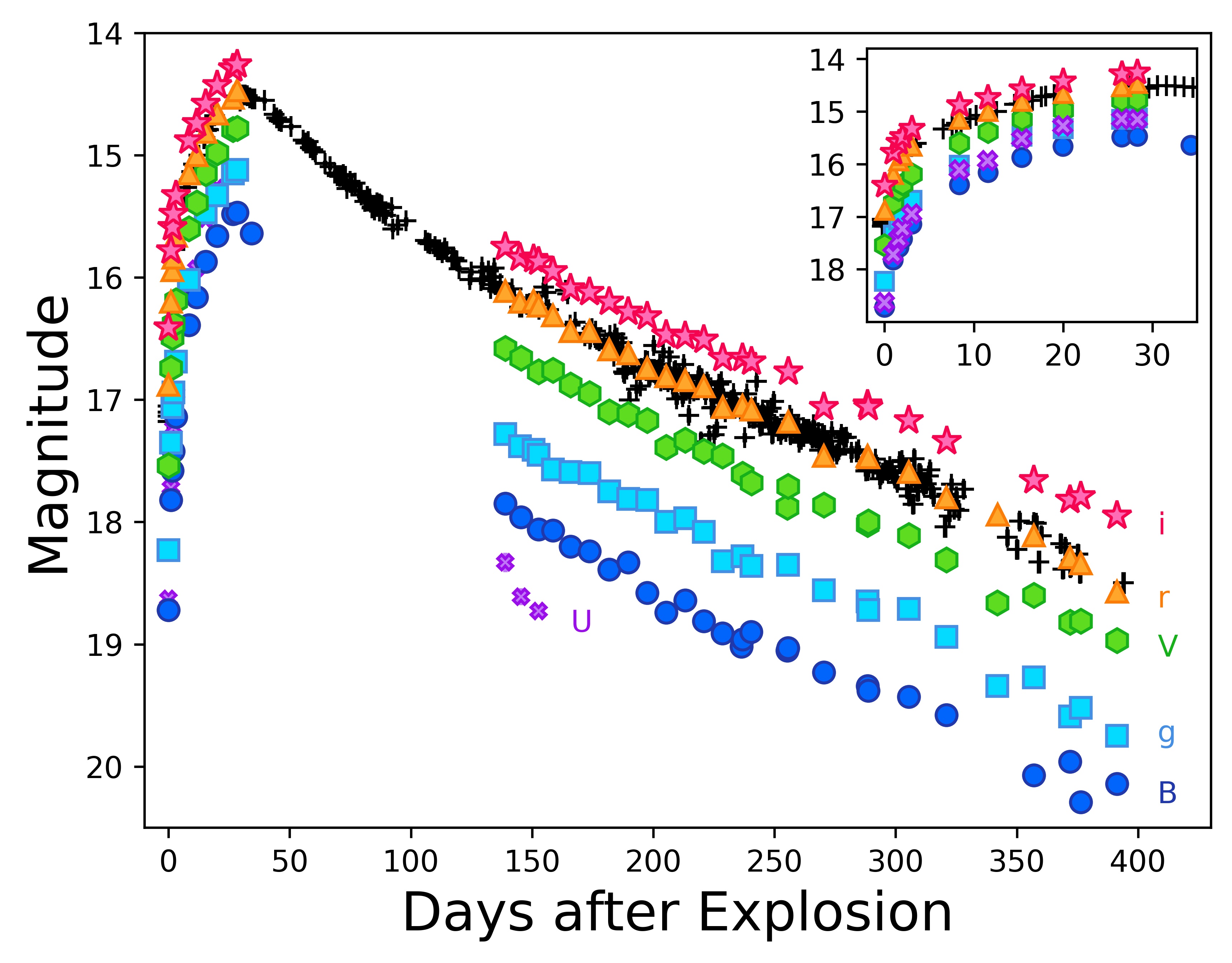}
\caption{Optical photometry of SN~2019esa.  The DLT40 $r$ data are shown as black plus marks and the marker size is larger than uncertainties for all data.   The adopted date of explosion is  MJD~58608.44 (2019 May 5.94). Inset shows a zoomed in region of the first 35 days. The dataset can be retrieved as the data behind the figure.}
\label{fig:fulllc}
\end{figure*}

A high-cadence photometric campaign by the Las Cumbres Observatory telescope network \citep{Brown_2013} was begun immediately after discovery, in the $UBVgri$ bands with the Sinistro cameras on the 1-m telescopes, through the Global Supernova Project. Using {\sc lcogtsnpipe} \citep{Valenti16}, a PyRAF-based photometric reduction pipeline, PSF fitting was performed. $UBV$-band data were calibrated to Vega magnitudes \citep{Stetson00} using standard fields observed on the same night by the same telescope. Finally, $gri$-band data were calibrated to AB magnitudes using the Sloan Digital Sky Survey \citep[SDSS,][]{sdssdr13}. The lightcurves are shown in Figure \ref{fig:fulllc}.

\begin{figure}
\includegraphics[width=\linewidth]{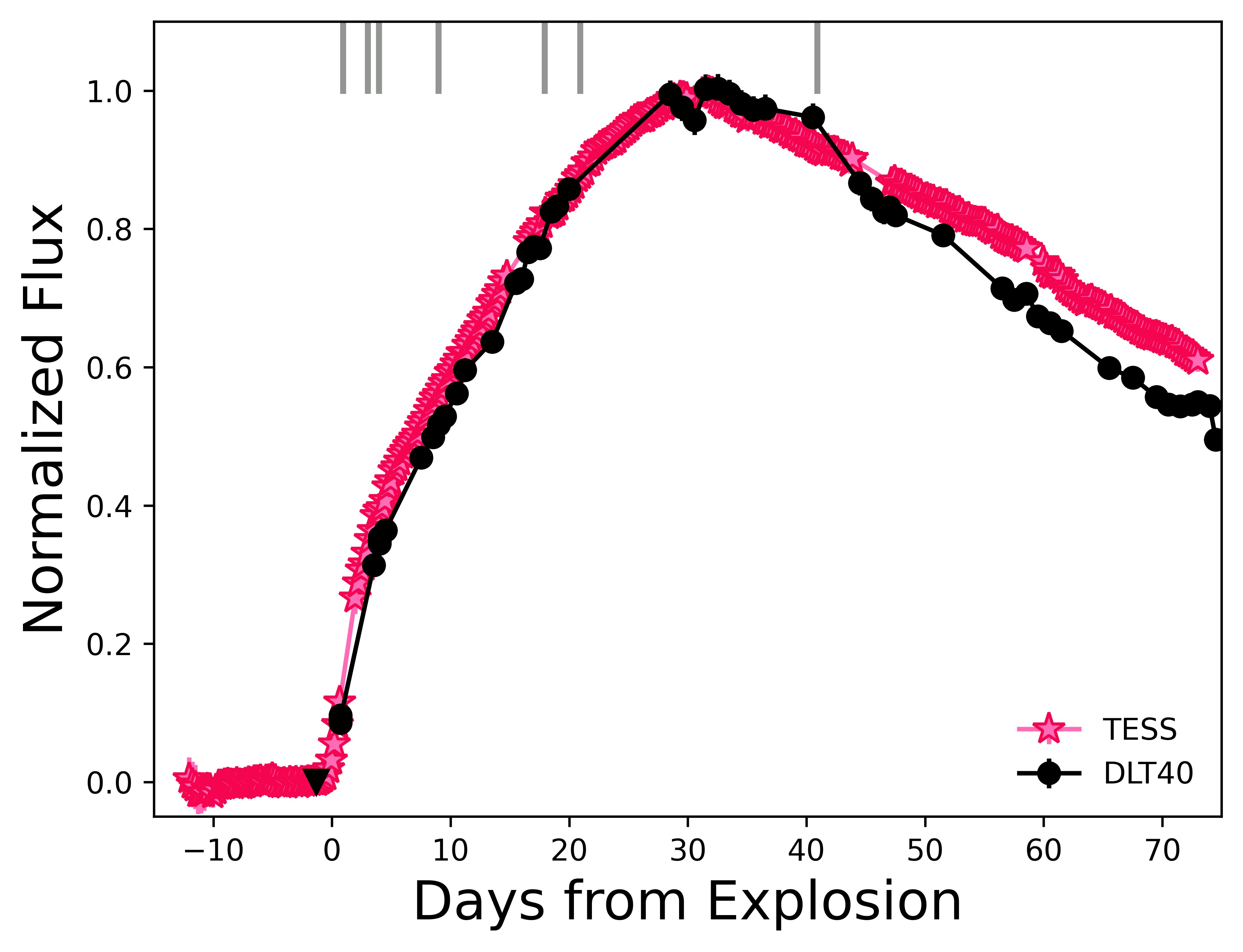}
\caption{TESS and DLT40 lightcurves of SN~2019esa normalized to max light. TESS points are a six hour rolling median, and the JD of explosion is taken to be 2458608.94. The last non-detection from DLT40 (JD 2458607.65) is shown as a black triangle, and epochs of FLOYDS spectroscopy are indicated at the top of the plot by gray lines. }
\label{fig:tess}
\end{figure}

SN 2019esa was also observed by the {\it Transiting Exoplanet Survey Satellite} \citep[{\it TESS},][]{Ricker2015} during the mission's Sector 11, 12, and 13 operations, from 2019-04-23 07:02:56.026 to 2019-07-17 20:01:19.027 UTC. The TESS lightcurve of SN~2019esa was previously published in \citet{2021MNRAS.500.5639V}, and we present it here as well. In order to reduce the inherent scatter, the data have been binned into 6-hour periods, beginning at the start of data acquisition. Where continuous monitoring was interrupted, the next 6-hour bin began when observations resumed. Bins with less than 5 data points were removed. Compromised epochs, as determined by \citet{2021MNRAS.500.5639V}, were excluded from this process. The median flux, maximum flux error, and the mean observation time of the 6-hour bin are shown in Figure \ref{fig:tess} plotted against the DLT40 photometry. Both datasets have been normalized to the maximum magnitude for ease of comparison. The deviation in the lightcurves after max between DLT40 and TESS is likely due to the redder transmission curve of the TESS filter.

\subsection{Spectroscopy}
\begin{figure*}
\includegraphics[width=7.1in]{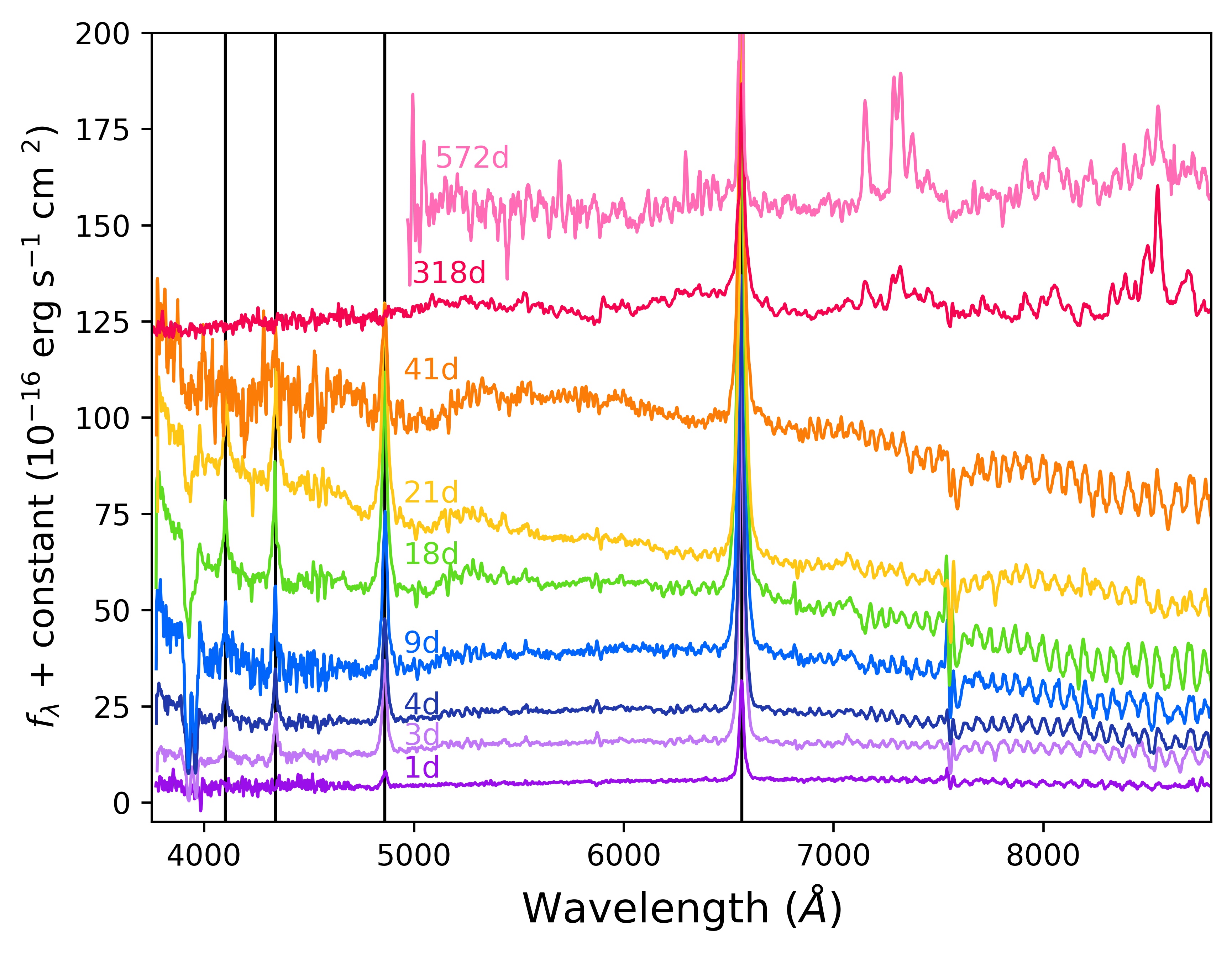}
\caption{Optical spectroscopic evolution of SN~2019esa listed in Table \ref{tab:optspec}. The hydrogen Balmer lines are indicated by vertical black line. The spectra have all been dereddened and the epochs are measured from the estimated date of explosion.  }
\label{fig:fullspec}
\end{figure*}

The majority of optical spectra were taken with the robotic FLOYDS spectrograph on the 2-m Faulkes Telescope South in Siding Springs, Australia through the Global Supernova Project \citep[FTS;][]{Brown_2013}. A 2$\arcsec$ slit was placed on the target at the parallactic angle. One-dimensional spectra were extracted, reduced, and calibrated following standard procedures using the FLOYDS pipeline \citep{Valenti14}. They are shown in Figure~\ref{fig:fullspec}. 

\input{Speclog}

One late-time spectrum was obtained with the Low Dispersion Survey Spectrograph 3 (LDSS-3) 
on the 6.5m Magellan Clay telescope at Las Campanas Observatory in Chile. Standard reductions were carried out using
{\sc iraf}
 including bias subtraction, flat fielding, cosmic ray rejection, local sky subtraction and extraction of one-dimensional spectra. The slit was aligned along the parallactic angle to minimize differential light losses, and flux calibration was done using a spectrophotometric standard taken that night at similar airmass. A log of the spectroscopic observations shown in Figure~\ref{fig:fullspec} can be found in Table~\ref{tab:optspec}.

\section{Light Curve Analysis} \label{sec:LC}

\subsection{Reddening Estimation}
The Milky Way line-of-sight reddening for ESO 035- G 018 is $E(B-V)_{MW} = 0.16$ mag \citep{2011ApJ...737..103S}.  No obvious \ion{Na}{1} D $\lambda\lambda$5889,5896 \AA\ absorption features are detected, preventing us from using measurements of the equivalent width (EW) of the lines to estimate the total reddening using the prescription of \citet{2012MNRAS.426.1465P}.   While the \ion{Na}{1} D absorption lines are not present,  from the early spectra it appears that the SN is considerably reddened, likely from CSM dust.   Therefore, we use comparisons with other well studied SN IIn to make an estimate of the amount of the total extinction towards SN~2019esa. 

\begin{figure}
    \centering
    \includegraphics[width=\linewidth]{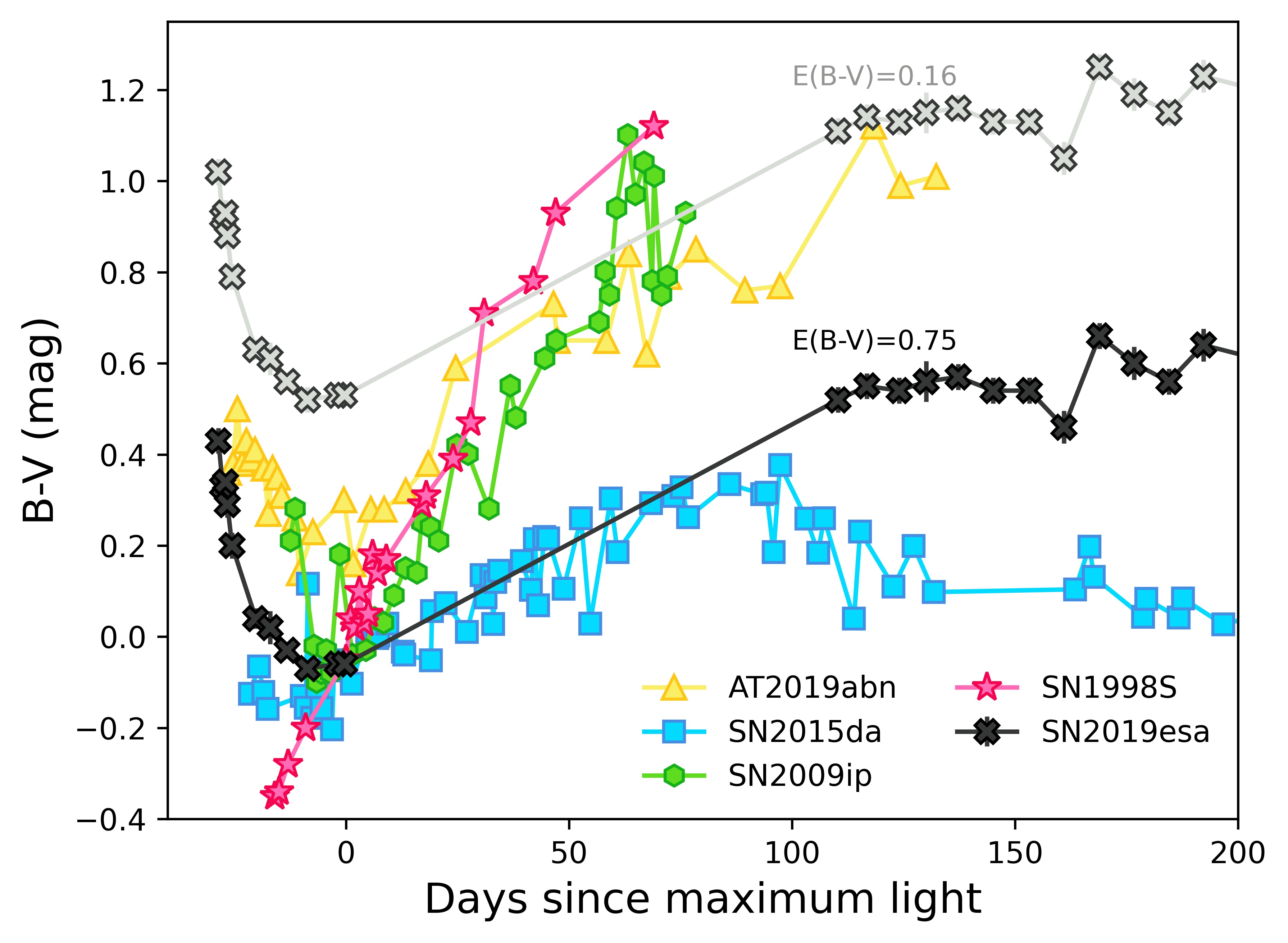}
    \includegraphics[width=\linewidth]{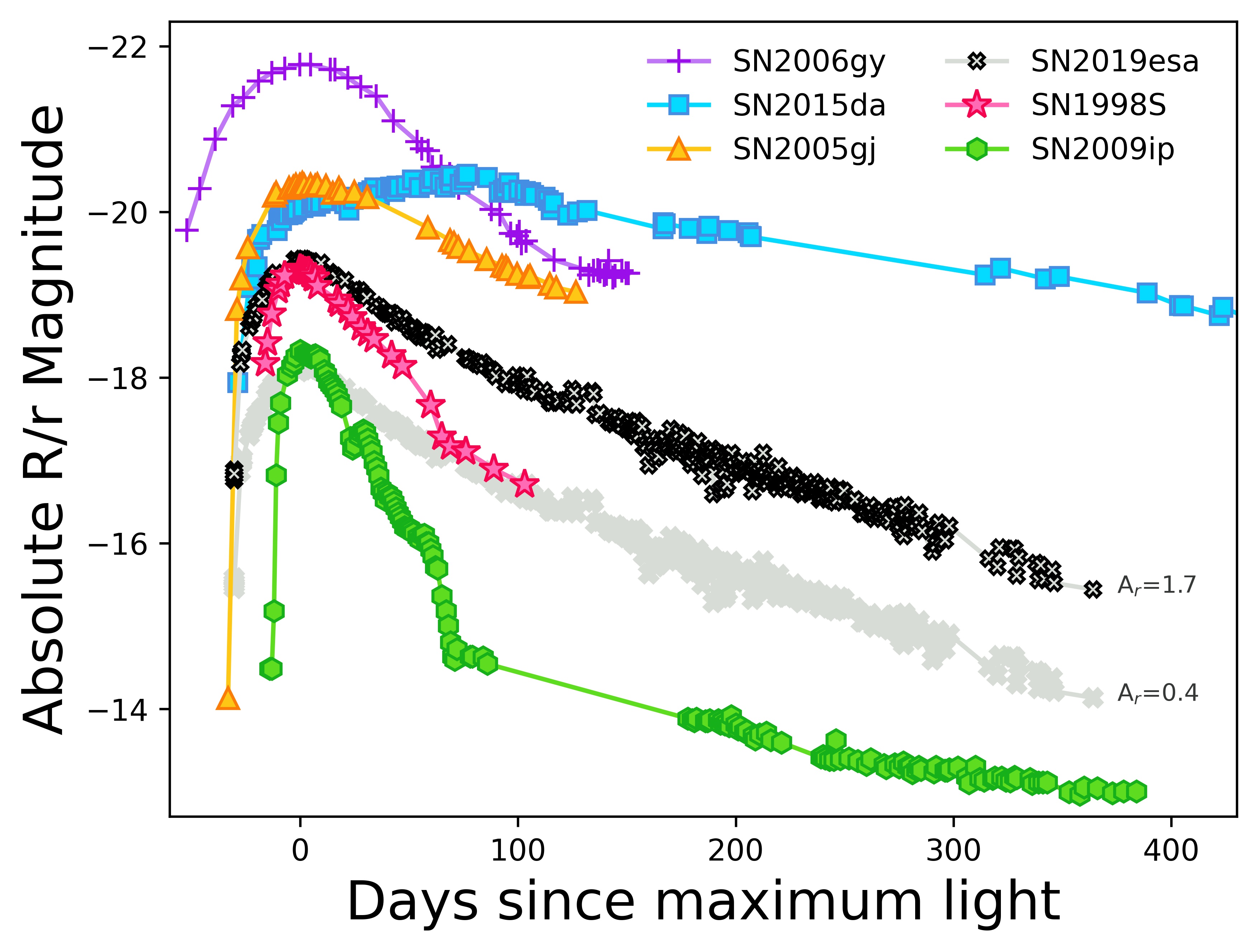}
    \caption{ $B-V$ color evolution (top) and absolute $r/R$ lightcurves (bottom) of SN~2019esa compared with other interacting SNe. Curves are shown for MW only extinction (gray) and $E(B-V) =$ 0.75 mag (black) for SN~2019esa. SN~1998S data from \citet{2000A&AS..144..219L}, SN~2005gj from \citet{2007arXiv0706.4088P}, SN~2009ip from \citet{2013ApJ...767....1P}, SN~2006gy from \citet{2007ApJ...666.1116S}, SN~2015da from \citet{2020A&A...635A..39T}, and AT2019abn from \citet{2020A&A...637A..20W}. SN~2015da has been shifted so that the inital rise matches that of SN~2019esa instead of the absolute maximum which was $\sim$100 days after explosion.}
    \label{fig:colorcompare}
\end{figure}

In the top panel of Figure \ref{fig:colorcompare} we show the $B-V$ color evolution of SN~2019esa along with a sample of other SN IIn which have been dereddened according to the published values. These include SN~2009ip \citep{2013ApJ...767....1P}, SN~1998S \citep{2000A&AS..144..219L}, and SN~2015da \citep{2020A&A...635A..39T}. Using the $B-V$ color at maximum brightness, we find that an $E(B-V)_{tot} = 0.75$ for SN~2019esa gives the best color match with SN~2009ip, SN~1998S, and SN~2015da. Converting this $E(B-V)$ value to an extinction using \citet{1999PASP..111...63F}, we obtain an A$_{r}$ = 1.72 mag.  When we apply this value to the  $M_r$ lightcurve (bottom Figure \ref{fig:colorcompare}) this also places SN~2019esa nicely among other interacting SNe.  

While we use $E(B-V)_{tot} = 0.75$ mag for the luminosity dependent values in this paper, we must caution that this is only our best estimate and by no means a definitive reddening value. Additionally, as we discuss below, it is also likely that the reddening may be variable during the first month of evolution as dust surrounding the SN is destroyed.

 \begin{figure*}
\includegraphics[width=3.3in]{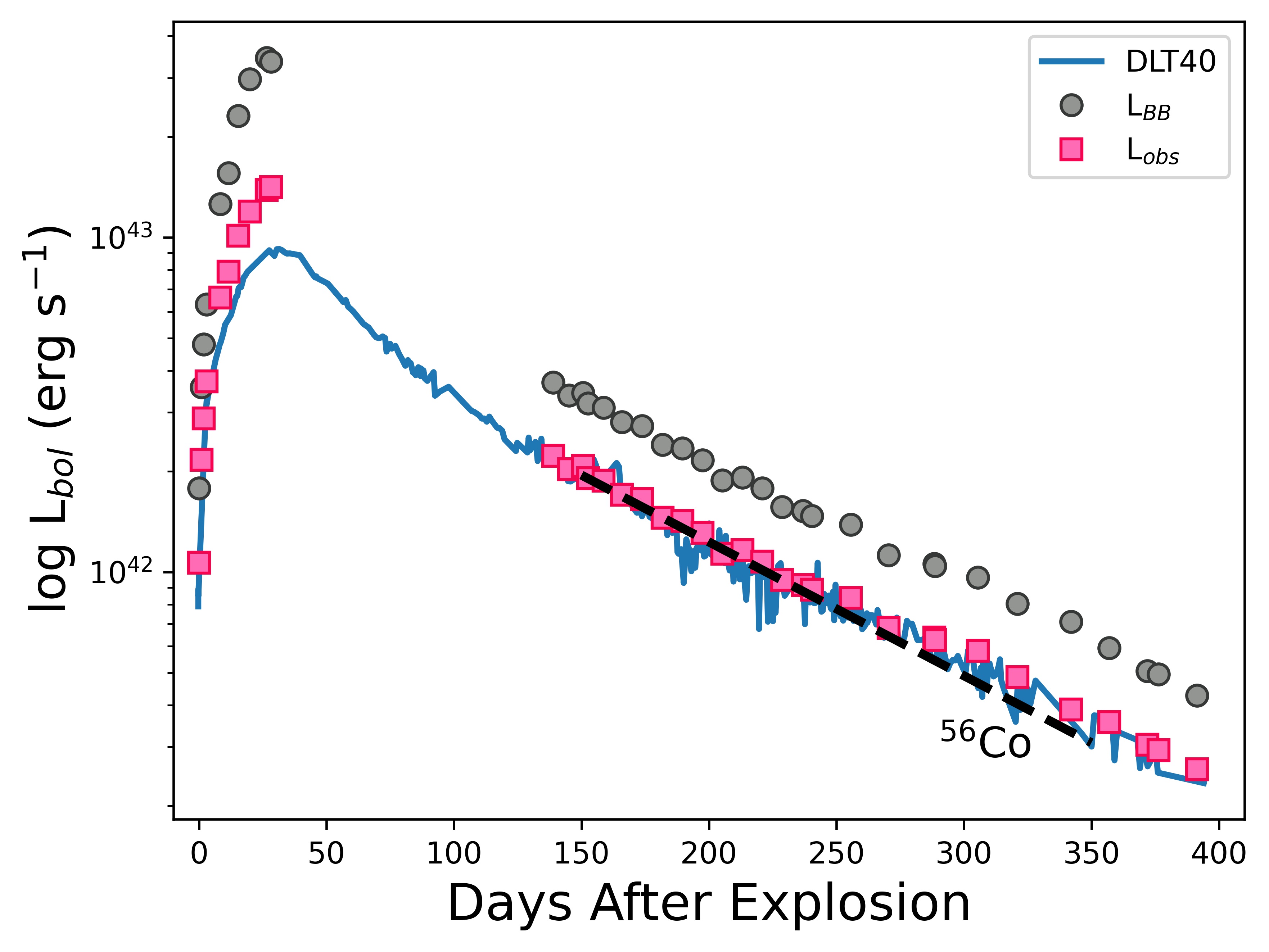}
\includegraphics[width=3.6in]{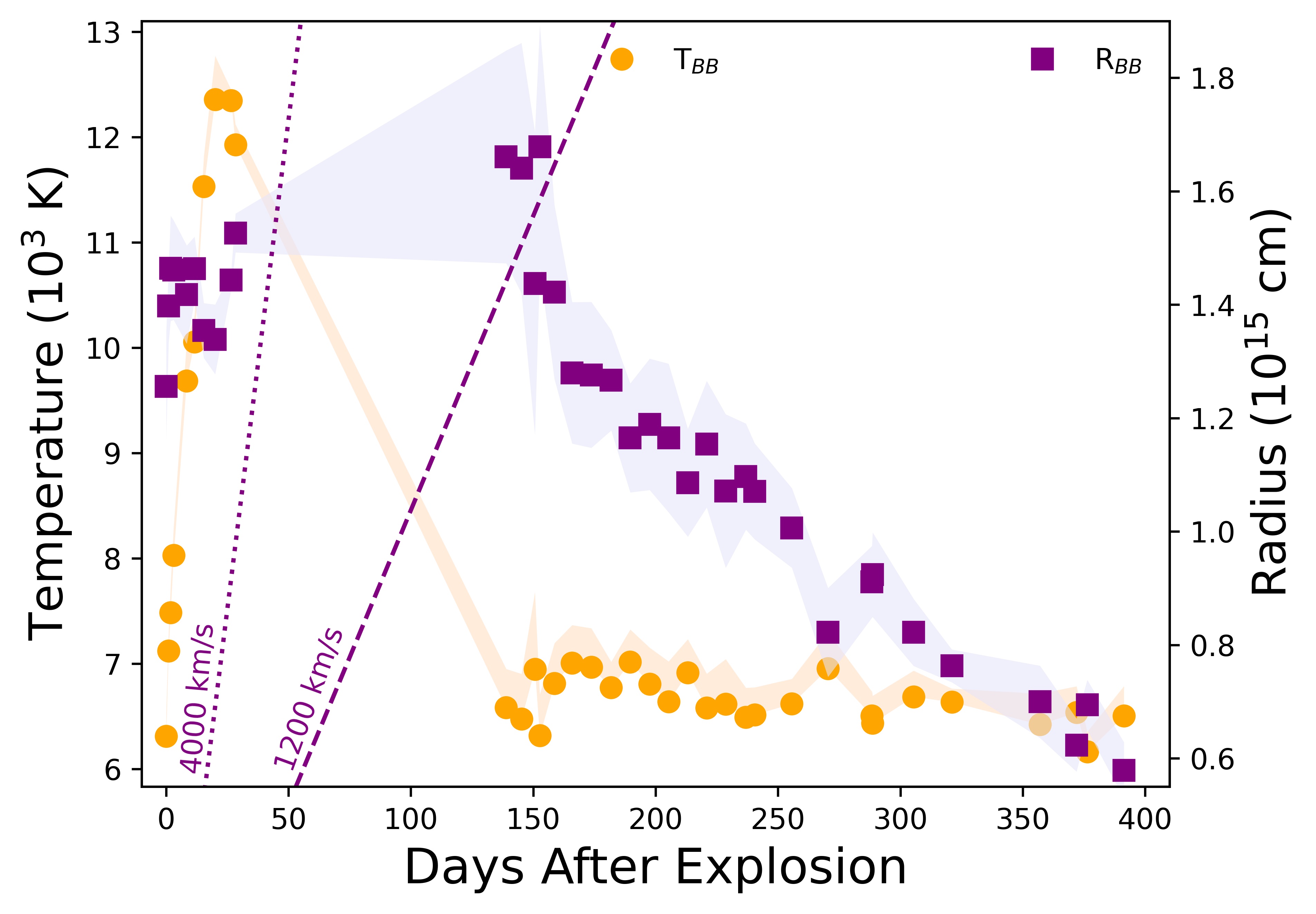}
\caption{Left: Pseudo-bolometric light curve of SN~2019esa constructed from the optical photometry.  The pink squares indicate the observed luminosity, while the gray circles come from blackbody corrected data. The $^{56}$Co decay rate is indicated by the dashed black line. The DLT40 lightcurve is also shown shifted to match the late-time observed bolometric luminosity to show the possible lightcurve evolution during the lack of Las Cumbres Observatory data. Right: Blackbody temperature and radius evolution of SN~2019esa derived from the optical photometry. The temperature is plotted in orange, and the radius in purple. The uncertainties are indicated by the shaded regions. Additionally lines of constant velocity of the FWZI (dotted) and FWHM (dashed) of H$\alpha$ are shown. All data have been dereddened by our assumed $E(B-V)_{tot}$ = 0.75.}
\label{fig:bolLC}
\end{figure*}

\subsection{Light curve evolution}
The full photometric evolution of SN~2019esa can be seen in Figure \ref{fig:fulllc}, including the high-cadence DLT40 $r$-band data. Following \citet{2021MNRAS.500.5639V} we use an explosion epoch of MJD 58608.44 derived from the TESS data. Both the DLT40 and TESS photometry indicate a date of maximum light of around MJD 58640, or roughly 32 days to a maximum $r$ = 14.51 mag. There is a break in the multi-band Las Cumbres Observatory coverage right around maximum light due to hour angle constraints, but the $B$-band data also shows a max around MJD 58636, or 28 days post-explosion.  For comparison, typical rise times for non-interacting Type II SNe are around 10 days in $r$-band \citep{2015MNRAS.451.2212G}, and roughly 19 days for a SN Ia \citep{2015MNRAS.446.3895F}. Introducing CSM interaction can lengthen the rise time depending on the radial density structure of the CSM, potentially increasing it to weeks or months in SNe IIn  \citep {2020A&A...637A..73N} and Ia-CSM \citep{2013ApJS..207....3S}.  

The relaxed hour angle constraints of the DLT40 monitoring allowed for continuous coverage out to roughly 400 days post explosion.  The resulting lightcurve shows an almost linear decline of 4 magnitudes from max until our last observation, or a rate of $\sim$ 0.011 mag day$^{-1}$. When Las Cumbres Observatory began observing again on MJD 58748 the $BgVi$ data all show a slightly slower decline rate of $\sim$ 0.009 mag day$^{-1}$, while the Las Cumbres Observatory $r$ is similar to that of DLT40 at $\sim$ 0.010 mag day$^{-1}$. This value is almost identical to that of the interacting SN~2005gj \citep{2007arXiv0706.4088P}, and similar to other values measured in other interacting SNe \citep{2012ApJ...744...10K,2016MNRAS.463.1088K},  although we expect a wide range of decline rates as the CSM around interacting SNe will have different densities and geometries.

The bottom panel of Figure \ref{fig:colorcompare} shows the absolute $r$ magnitude lightcurve of SN~2019esa compared to the $r/R$ band lightcurves other SNe IIn and Ia-CSM/IIn. All SNe have been dereddened by values listed in the literature. Because of the uncertainty in the reddening of SN~2019esa we show two lightcurves, one using the Milky Way only extinction value of A$_{r}$ = 0.4 mag, and the other using the value derived from the color comparison of A$_{r}$ = 1.7 mag.  This puts the range of absolute magnitude at maximum between --18.1 $<$ M$_{r}$ $<$ --19.4 mag. These are quite similar to the max M$_{r}$ for IIn SNe SN~2009ip and SN~1998S, respectively, but significantly fainter than the IIn SN~2015da and the SLSN SN~2006gy. We note that when we deredden the spectra of SN~2019esa so that the absolute magnitude at peak matches that of SN~2006gy (A$_{r}$ = 4.1 mag) the SED of SN~2019esa becomes too blue to be physically possible.
 
 While the absolute magnitude at maximum of SN~2019esa may be similar to SN~2009ip and SN~1998S, the rise time is significantly slower. Comparatively, the $\sim$30 day rise to max is much shorter than the rise time for SN~2006gy which took about 70 days to reach an absolute magnitude of $-$21 \citep{2007ApJ...666.1116S}, or the IIn SN~2015da which had a very quick rise in brightness but then took around 100 days to reach the absolute maximum \citep{2020A&A...635A..39T}. The potential Ia-CSM SN~2005gj $r$-band rise was almost identical to SN~2019esa at 31.7 days \citep{2007arXiv0706.4088P}, although the bluer bands of SN~2005gj rose much faster (12.7 days in $u$, 18.5 in $g$), while all optical colors of SN~2019esa appear to have similar $\sim$30 day rises.

\subsection{Color, Temperature, and Radius Evolution}
Unlike most IIn SNe, SN~2019esa has an unusual color evolution (top panel, Figure \ref{fig:colorcompare}) where it becomes bluer as the lightcurve rises to peak, then slowly reddens after max brightness. While uncommon for most CCSNe, this behavior has been seen to varying amounts in the IIn SN~1988Z \citep{1993MNRAS.262..128T}, SN~2006aa \citep{2013A&A...555A..10T}, and in SN~2009ip \citep{2013MNRAS.433.1312F,Graham14}.  It has also been noted in the ILRT objects AT~2017be \citep{2018MNRAS.480.3424C}, SNHunt120 \citep{2020A&A...639A.103S}, and AT~2019abn \citep{2019ApJ...880L..20J}.   In the case of AT~2019abn, \citet{2019ApJ...880L..20J} suggest the early evolution to bluer colors may be due to continuous dust destruction in the CSM surrounding the star.   We surmise that this could be at least partly responsible for the odd color evolution in SN~2019esa.

In the left panel of Figure \ref{fig:bolLC} we show a  quasi-bolometric lightcurve of SN~2019esa created using the Light Curve Fitting package from \citet{2020zndo...4312178H}. Similarly to what was done in \citet{2022arXiv220308155H}, a blackbody spectrum was fit to the SED at each epoch, and the resulting best fit blackbody was integrated over the available bands to create a pseudobolometric lightcurve.  The resultant temperature and luminosity from the SED fits are also shown in the right panel of Figure \ref{fig:bolLC}. The data have been corrected for an $E(B-V)_{Tot} = 0.75$ mag and the distance modulus $\mu$ = 32.22 mag. The bolometric lightcurve produced only from the observations are shown in pink, while the blackbody corrected  bolometric lightcurve is shown in gray. We have also plotted the DLT40 open filter lightcurve in blue, shifted to match the late-time observed bolometric lightcurve, to show the likely shape of the L$_{bol}$ during the gap in the Las Cumbres Observatory coverage. Integrating over the blackbody corrected lightcurves for the first 400 days (using a straight line interpolation over the gap) gives a total radiated energy of $E_{\rm rad} \approx 1 \times 10^{50}$ ergs.
This is likely a lower limit though, as no UV or IR photometry was used in deriving the bolometric luminosity, and the total reddening could be even greater than the  value used.

From the bolometric lightcurves we can see that between days $\sim$125--250 the luminosity decline is similar to that of fully-trapped $^{56}$Co decay of 0.98 mag 100 d$^{-1}$, but may deviate by small amounts at later times.  If we estimate the mass of $^{56}$Ni at 200 days using L$_{bol}$ we obtain a value of M$_{^{56}Ni}$ = 0.30 M$_{\sun}$ \citep{2015ApJ...806..225P}. Of course this is likely an overestimate since the CSM interaction could still be strong and contributing to the late-time luminosity, although without adequate spectroscopic coverage during this time it is difficult to determine the amount of contribution. For comparison, normal Type II SNe have an average M$_{^{56}Ni}$ = 0.04 M$_{\sun}$ \citep{Valenti16,2017ApJ...841..127M,2021MNRAS.505.1742R}, and type Ia SNe 0.1 M$_{\sun}$ $<$ M$_{^{56}Ni}$ $<$ 1.1M$_{\sun}$ \citep{1997A&A...328..203C}.

Just as in the color evolution of SN~2019esa, the evolution of T$_{BB}$ and R$_{BB}$, shown in the right panel of Figure \ref{fig:bolLC}, is also very unusual. Unlike most SN IIn which have their highest temperature at explosion (or at least at discovery) and cool quickly with time \citep{2013A&A...555A..10T}, T$_{BB}$ for SN~2019esa actually rises over the first $\sim$35 days from a relatively cool $\sim$6000 K to $\sim$13000 K. This temperature then drops at some point during our lack of multi-wavelength coverage, but then stays relatively constant between 6000--7000 K from day 125 until day 400. This low, late-time T$_{BB}$ of $\sim$6500 K is commonly seen in IIn SNe. The range in T$_{BB}$ seen in SN~2019esa is similar to the interacting SNe 2006gy, 2005gj, and 1998S \citep{2010ApJ...709..856S,2000MNRAS.318.1093F,2007arXiv0706.4088P}, although these objects do not show the 30 day temperature rise after explosion.  This strange behavior can be understood  by also looking at the color and spectral evolution.  As the luminosity rises over the first few months the flux rises more in the blue than in the red, which could be due to dust destruction in the surrounding CSM.  Therefore it is possible that there was much less extinction near maximum light than at explosion, underestimating the early time temperatures and giving the appearance that the temperatures are actually rising when there could just be variable extinction. 

The evolution of the R$_{BB}$ of SN~2019esa rises to a maximum value around 150 days after explosion, only to fall again over the next 250 days. It is important to emphasize here that in SN IIn, R$_{BB}$ does not carry much importance since the interaction often removes much of the physical meaning. The narrow range of R$_{BB}$ values between 0.6 -- 1.8 $\times$ 10$^{15}$ cm derived for SN~2019esa is consistent with other IIn SNe \citep{2013A&A...555A..10T}, but smaller than the much brighter interacting SN~2006gy and SN~2005gj whose R$_{BB}$ had maximum values of 4 and 8  $\times$ 10$^{15}$ cm respectively \citep{2010ApJ...709..856S,2007arXiv0706.4088P}.
Interestingly, the peak of R$_{BB}$ at around 150 days is one of the latest seen in interacting SNe, much later than the $\sim$50 days for SNe 1998S \citep{2000MNRAS.318.1093F} and 2005gj \citep{2007arXiv0706.4088P} and slightly longer than the $\sim$115 days for SN~2006gy \citep{2010ApJ...709..856S}. We also plot in the right panel of Figure \ref{fig:bolLC} the constant velocity curves of material moving at 4000 km s$^{-1}$ and 1200 km s$^{-1}$, the FWZI and the FWHM respectively of the H$\alpha$ emission shown in Figure \ref{fig:Hacomp}.

We note that the turnover in R$_{BB}$ occurs at roughly the intersection of a constant 1200 km s$^{-1}$ velocity, the FWHM of the intermediate-width line. The epoch when R$_{BB}$ stops increasing, either by stalling at a constant value or even decreasing as we see here, may occur when the forward shock has finally traversed the majority of the CSM. Unfortunately we have no optical spectra during this time to check for any obvious physical changes that would support this claim. If we assume that it does take 150 days to traverse the bulk of the CSM, at a velocity of 1200 km s$^{-1}$, this suggests the circumstellar shell extends to about 1.5 $\times$ 10$^{15}$ cm.  If the surrounding CSM material is similar to that of SN~2006gy or SN~2006tf, R$_{BB}$ does not represent the true photospheric radius and the decrease in R$_{BB}$ may be due instead to the shell covering factor decreasing as the optical depth drops \citep{2007ApJ...671L..17S,2008ApJ...686..467S}. In other words, the post-shock shell may be very clumpy, allowing for regions of high optical depth dispersed within an optically thin medium.

\section{Spectral Evolution} \label{sec:specev}

SN~2019esa shows spectroscopic characteristics typical of SN IIn, with strong, intermediate-width ($\sim$1000 km s$^{-1}$) Balmer emission. While normally early spectra of Type II SNe (interacting or not) show very blue continua that are either mostly featureless, or exhibit high-ionization emission lines such as 
\ion{N}{5} $\lambda\lambda$4434,4641, \ion{He}{2} $\lambda$4686, and \ion{C}{4} $\lambda\lambda$5801,5812, SN~2019esa appears quite red with only strong hydrogen emission lines and deep Ca H \& K absorption.   A dense, hydrogen rich, CSM enshrouding the SN could act to dampen the strengths of any of the high-ionization features.

Over the first two weeks the spectra do gradually become bluer as the luminosity and temperature continues to rise. The strong Balmer emission lines seen in the spectra can be easily fit by a single Lorentzian profile for all but the last epoch. In the last epoch the Lorentzian profile has disappeared and H$\alpha$ can best be fit by a Gaussian with a FWHM of 900 km s$^{-1}$. Table \ref{tab:FWHM} lists the measured FWHM, fluxes, and luminosities of H$\alpha$ and H$\beta$ only, but note the bluer Balmer emission lines show similar widths.  The Lorentzian profile is likely caused by electron scattering in the optically thick CSM \citep{2001MNRAS.326.1448C,2008ApJ...686..467S}, therefore the FWHM may be a slight overestimate of the true shock velocity if electron scattering is present in the wings of the lines. This seems to still be occurring as late as 318 days after explosion. Combining the persistence of electron scattering along with the rouhgly constant FWHM of the lines (H$\alpha$ in particular) for the first year after explosion, it is clear that the CSM surrounding SN~2019esa must be incredibly dense.

\begin{figure}
\includegraphics[width=\linewidth]{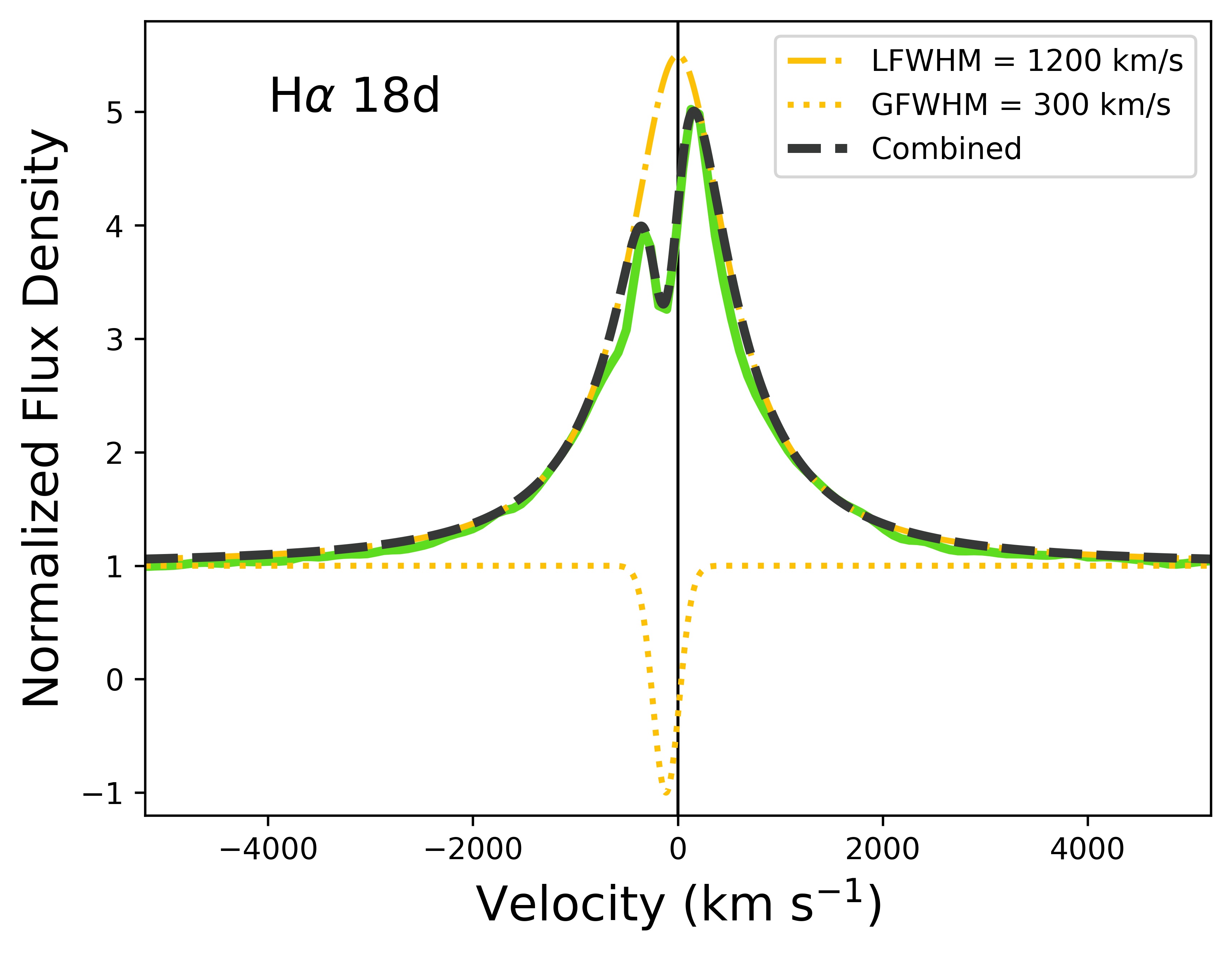}
\includegraphics[width=\linewidth]{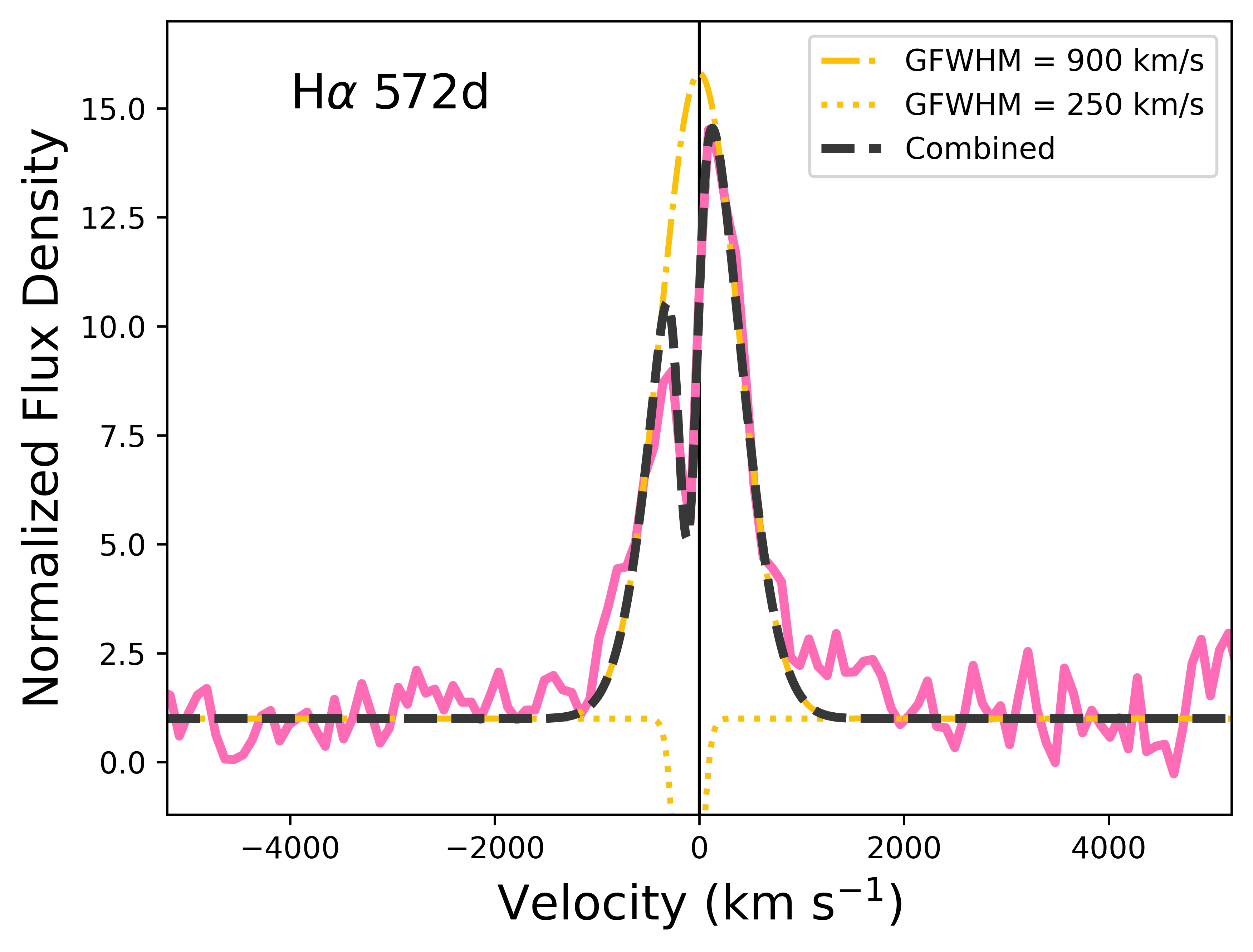}
\caption{The H$\alpha$ emission line on day 18 (top) and 572 (bottom).  Day 18 can be reproduced by a Lorentzian  emission profile combined with a much narrower Gaussian absorption profile.  This creates a P Cygni feature with a FWHM of 300 km s$^{-1}$. Day 572 is better represented by a 900 km s$^{-1}$ Gaussian emission profile combined with a 250 km s$^{-1}$ absorption profile.  }
\label{fig:Hacomp}
\end{figure}

In the day 18 spectrum, which had the best seeing of all of the FLOYDS spectra, a narrow H$\alpha$ absorption component is seen at $-$115 km s$^{-1}$ (Figure \ref{fig:Hacomp}, top). A similar feature is also seen at the same location in the higher resolution, day 572 LDSS3 spectrum (Figure \ref{fig:Hacomp}, bottom).  The absorption line is too close to H$\alpha$ to be from over-subtraction of galactic [\ion{N}{2}] lines, and its appearance in two different epochs with two different instruments suggests that it is real.  The FWHM of the absorption line ranges between 250--300 km s$^{-1}$, velocities consistent with the winds of LBVs and BSGs, although the low resolution of the instruments cannot rule out slower RSG or YSG progenitor systems. The persistence of this narrow emission line in our last spectrum also suggests the existence of CSM which has yet to be traversed by the SN shock.

Unlike most interacting SNe, at no time do we see any underlying broad emission from the quickly expanding SN ejecta, the caveat being there are large gaps in the spectral coverage after the first two months, and in this gap during the decline from peak is when one would expect to see these broad lines appear. This indicates that the continuum emission coming from the CSM interaction is (at least at early times) producing a photosphere outside of the forward shock,  prohibiting a glimpse into the internal SN ejecta.  At late times when the CSM interaction region has expanded out to larger radii containing less dense pre-shock gas it finally becomes optically thin, but the SN ejecta has faded significantly so that it becomes non-detectable over the much stronger CSM interaction.

\begin{figure*}
 \includegraphics[width=3.5in]{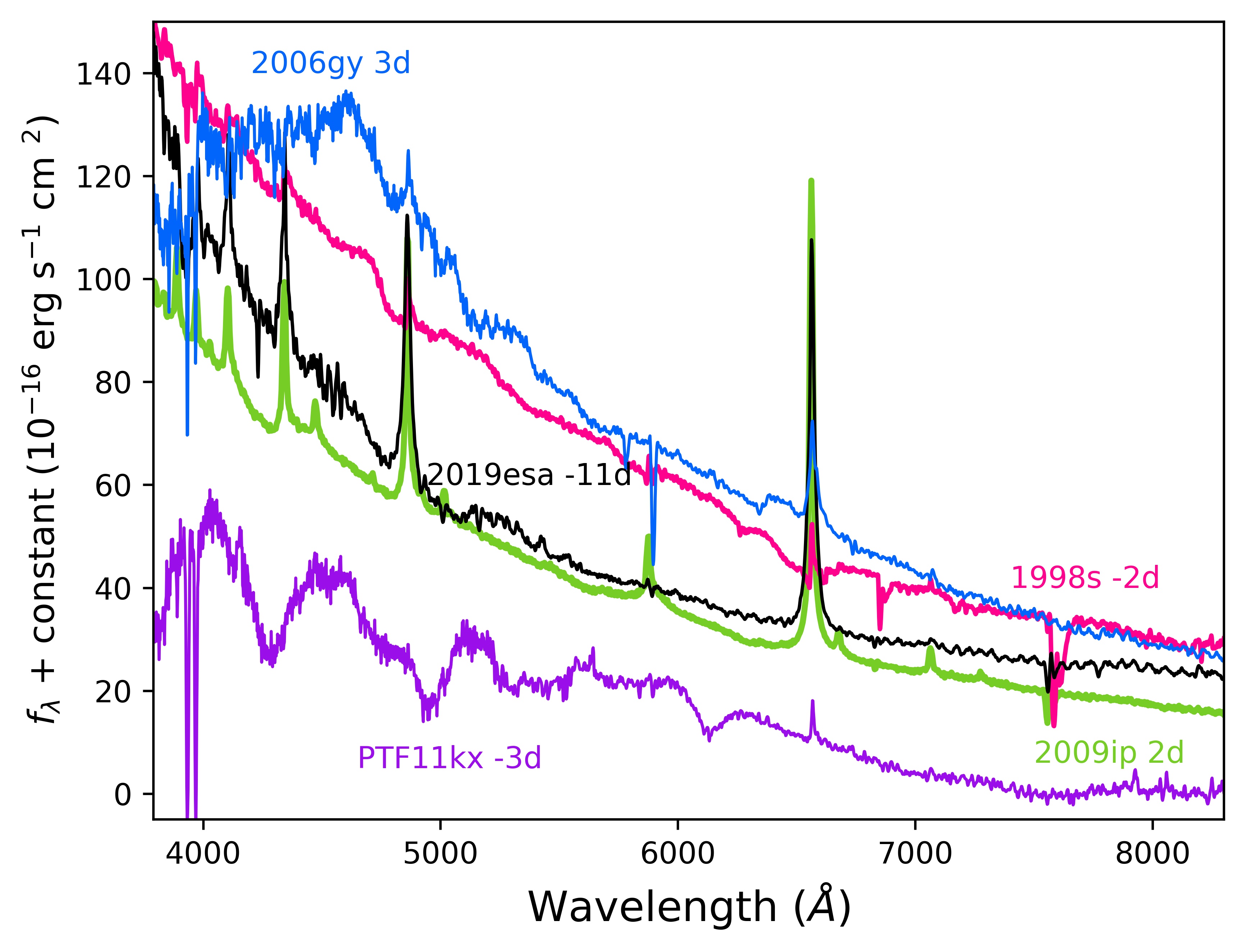}
\includegraphics[width=3.6in]{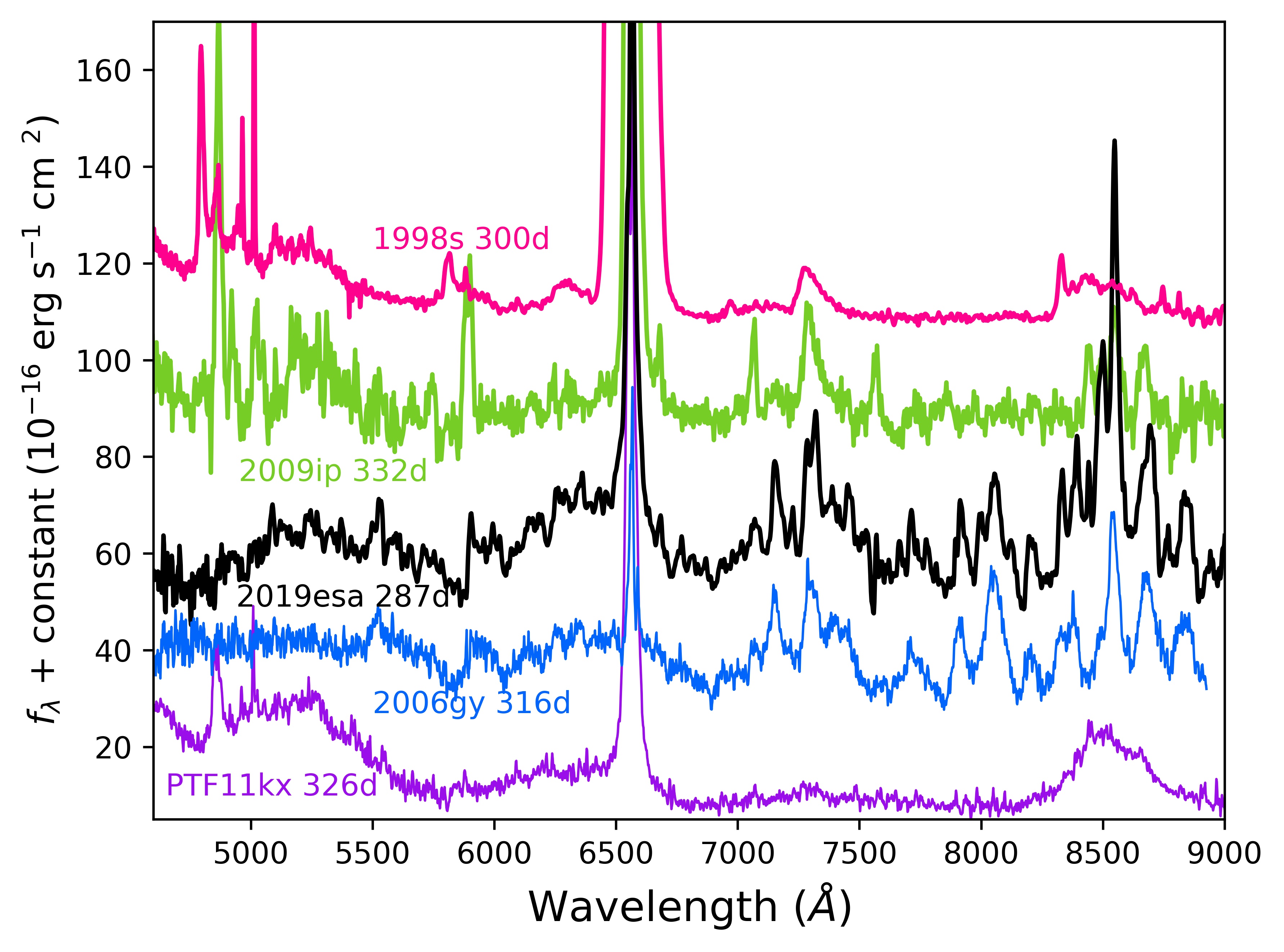}
\caption{Left: Comparison of optical spectra of SN2019esa (black) with other IIn and Ia-CSM SNe near maximum light. Spectra have been corrected for extinction and come from  \citet[SN~2006gy]{2007ApJ...666.1116S}, \citet[SN~1998S]{2001MNRAS.325..907F}, \citet[SN~2009ip]{2013MNRAS.433.1312F}, and \citet[PTF11kx]{2012Sci...337..942D}.  The epoch with respect to maximum brightness is listed by the SN name. Right: Same as left figure, but at later times. Spectra are from \citet[SN~1998S]{2000ApJ...536..239L}, \citet[SN~2009ip]{2014MNRAS.438.1191S}, \citet[SN~2006gy]{2009ApJ...697..747K}, and \citet[PTF11kx]{2013ApJ...772..125S}.}
\label{fig:spectracompare}
\end{figure*}

In the last two epochs it is clear that the spectra are still strongly dominated by CSM interaction, and lack some of the normal nebular emission lines that are usually present in typical CCSNe without CSM interaction, mainly O and He. The lack of [\ion{O}{1}] is not surprising since it is rarely seen in interaction powered supernovae \citep{2015MNRAS.447..772F}, but \ion{He}{1} is often seen in SN IIn (for instance SN~2009ip and SN~1998S, shown in Figure \ref{fig:spectracompare}). It is likely that we are beginning to see some of the cool dense shell (CDS)  due to the emergence of the emission lines of [\ion{Ca}{2}], the \ion{Ca}{2} IR triplet, and possibly a Paschen 10-3 line at 9017 \AA\ . Most intriguingly are the presence of possible neutral Fe lines, which we will discuss in more detail below. Although there is significant blending, the widths of the lines are roughly 1000 km s$^{-1}$, consistent with that of H$\alpha$.  This suggests that these lines form from the same location as the hydrogen lines, likely in the post shock CDS.

\begin{table}
 \begin{center}
\caption{Balmer line properties}
\begin{tabular}{@{}lcccccc}\hline\hline
 Age&
\multicolumn{3}{c}{ H$\alpha$\tablenotemark{a,}\tablenotemark{b,}\tablenotemark{c}}&
\multicolumn{3}{c}{H$\beta$}\\
days &  FWHM &Flux& Log(L)&FWHM& Flux& Log(L) \\
\hline
1&1020& 3.8 & 40.54&1600& 1.1&40.0\\
3&1100&15.3&41.15&1450&6.5&40.78\\
4&1115&18.3&41.23&1515&7.3&40.85\\
9&1190&29.6&41.45&1575&12.0&41.0\\
18&1200&31.6&41.46&1610&16.6&41.18\\
21&1200&26.8&41.40&1840&18.8&41.23\\
41&1295&22.6&41.32&1960&8.6&40.90\\
318&1320&3.3&40.48&--&--&--\\
572&900\tablenotemark{d}&0.02&38.26&--&--&--\\
\hline

\hline
\end{tabular}\label{tab:FWHM}
\tablenotetext{a}{FWHM units are km s$^{-1}$.}
\tablenotetext{b}{Flux units are in 10$^{-13}$ erg s$^{-1}$ cm$^{2}$.}
\tablenotetext{c}{Luminosity units are in erg s$^{-1}$.}
\tablenotetext{d}{Gaussian.}
\end{center}
\end{table}

\section{Discussion} \label{sec:Disc}

\subsection{Location and size of the CSM}

The fact that no broad ejecta lines are seen in our spectra, and the persistence of intermediate-width lines, even in our earliest spectra, suggest that the CSM surrounding SN~2019esa is quite dense.   Similar to the very luminous IIn SN~2006tf, it is likely that the SN shock is quickly decelerated by the massive CSM, preventing the fast ejecta expansion that gives rise to the broad emission lines and instead converted the bulk of the kinetic energy into thermal energy and light \citep{2008ApJ...686..467S}.  This would also explain the slower ($\sim$ 30 day) rise to maximum of the lightcurve as the energy deposited into the optically thick CSM would have to slowly diffuse out \citep{2007ApJ...671L..17S}. Additionally, the narrow P-Cgyni absorption is still seen in our last spectrum on day 572, indicating the presence still of some slow, unshocked CSM which would be at a radius of at least 6 $\times$ 10$^{15}$ cm.

Although we do not have any X-ray or radio data of SN~2019esa, we can use the luminosity and kinematics of the system offered to us through our bolometric lightcurve and optical spectra to determine a wind-density parameter $w$ and therefore a rough estimate of the mass-loss rate of the progenitor star.  This can be expressed as $w = \frac{\dot{M}}{V_{CSM}}$ = $\frac{2L}{V_{SN}^{3}}$, where V$_{CSM}$ is the CSM velocity measured from the minimum of the narrow P-Cygni lines and V$_{SN}$ is the supernova expansion velocity \citep[see][]{2017hsn..book..403S}.   At early times the speed of the shock is unknown because it is hidden below the photosphere, but from our day 318 spectrum, V$_{SN}$ = 1320 km s$^{-1}$ (taken from the FWHM of H$\alpha$)
and we assume  V$_{CSM}$ = 300 km s$^{-1}$ from our day 18 and day 572 spectra (although this is likely an upper limit).  From our bolometric lightcurve we measure L = 8 $\times$ 10$^{41}$ erg s$^{-1}$ on day 318, resulting in an $\dot{M}$ = 0.3 M$_{\sun}$ yr$^{-1}$. If we assume that the pre-SN mass loss occurred 3-4 years prior ($ t = 318d \times V_{SN}/V_{CSM} $), this would suggest a total mass loss of 1-2 M$_{\sun}$.

Mass loss rates of a sample of IIn from \citet{2012ApJ...744...10K} range between  0.026--0.12 M$_{\sun}$ yr$^{-1}$, and in particular  $\dot{M}$ $\sim$ 2 $\times$ 10$^{-5}$ M$_{\sun}$ yr$^{-1}$  for SN~1998S \citep{2001MNRAS.325..907F}, and $\dot{M}$ $\sim$ 1--6 $\times$ 10$^{-5}$ M$_{\sun}$ yr$^{-1}$ for the interacting SN~2005gj \citep{2008A&A...483L..47T}. The extremely long-lived IIn SN~2015da did have a very high $\dot{M}$ = 0.6--0.7 M$_{\sun}$ yr$^{-1}$ \citep{2020A&A...635A..39T}, while mass loss rates of SN~2006gy range anywhere between  0.1 -- 0.5 M$_{\sun}$ yr$^{-1}$ \citep{2007ApJ...666.1116S,2013MNRAS.428.1020M}. The only progenitor channel capable of mass loss rates on the order of 0.5 M$_{\sun}$ yr$^{-1}$, would be from a prior eruption of an LBV, akin to the Great Eruption of $\eta$ Car \citep{2003AJ....125.1458S}.
 The 1-2 M$_{\sun}$ of CSM we estimate for SN~2019esa could have been lost over a short amount of time during a similar eruptive event. 

\subsection{Comparison to SN~2006gy}

SN~2006gy was a SLSN maintaining a magnitude of at least $-$21 mag for well over 100 days, with  Type IIn spectra \citep{2007ApJ...659L..13O,2009ApJ...691.1348A,2007ApJ...666.1116S,2010ApJ...709..856S}.  While SN~2019esa is less luminous, there are similarities between the optical spectra and the shape of the lightcurve of the two objects that are worth exploring. 

In the lower panel of Figure \ref{fig:colorcompare} we show the absolute r/R-band lightcurves of both objects.  The differences in peak magnitudes are easily seen, with the total radiated energy of SN~2006gy estimated to be $\sim$ 2 $\times$ 10$^{51}$ ergs \citep{2010ApJ...709..856S}, while SN~2019esa appears  less energetic at $\sim$ 1 $\times$ 10$^{50}$ ergs.  The rise times are also noticeably different, with the substantially more luminous SN~2006gy taking roughly twice the amount of time as SN~2019esa to rise to max at $\sim$70 days \citep{2007ApJ...666.1116S}. Various estimates from observations and models for SN~2006gy all require a high CSM mass of order 10-25 $M_{\odot}$ \citep{2007Natur.450..390W,2007ApJ...671L..17S,2007ApJ...666.1116S,2010ApJ...709..856S,2013MNRAS.428.1020M}. The SN ejecta mass must have at least a comparable mass in order to maintain the high observed speed of SN~2006gy's CDS \citep{2007Natur.450..390W,2010ApJ...709..856S}.  A mass loss of such a high amount plus similarly massive ejecta  requires an extremely massive progenitor star. A similar scenario, with a less massive CSM of 1--2 $M_{\odot}$ may be responsible for the observational characteristics of SN~2019esa.

At all comparable epochs there is a lack of \ion{He}{1} emission, unlike in some other IIn SNe (for example SN~2009ip shown in Figure \ref{fig:colorcompare}), which is consistent with the low ($<$ 15000 K) temperatures of both objects.  In the SN~2006gy spectrum taken at roughly 316 days after maximum, H$\alpha$ is extremely weak while it appears to be quite strong in SN~2019esa around the same age. In fact, comparison of the H$\alpha$ luminosities of SN~2006gy presented in \citet{2010ApJ...709..856S} with those of SN~2019esa listed in Table \ref{tab:FWHM} shows that the H$\alpha$ brightness of SN~2019esa is roughly 4$\times$ brighter than that of SN~2006gy at all comparable phases. Other than this one difference, it is these late-time spectra that show the most similarities between the two objects, specifically the intermediate-width, resolved \ion{Fe}{1}, \ion{Fe}{2}, and \ion{Ca}{2} emission lines seen redward of H$\alpha$.  

There has been some disagreement in the literature as to whether SN~2006gy was a core collapse or thermonuclear event, although there is consensus that interaction with massive CSM did occur.  Late-time HST and Keck AO imaging eliminate the need for a pulsational pair instability model for the explosion of SN~2006gy after revealing the presence of IR and scattered light echoes \citep{2008ApJ...686..467S,2010AJ....139.2218M,2015MNRAS.454.4366F}, and an interaction powered model with up to 10--25 M$_{\sun}$ of CSM is required to reproduce the high luminosity and long duration of the light curve  \citep{2007Natur.450..390W,2007ApJ...671L..17S,2007ApJ...666.1116S,2009ApJ...691.1348A,2010ApJ...709..856S,2013MNRAS.428.1020M}. More recently, though, \citet{2020Sci...367..415J} analysed the late-time spectra of SN~2006gy, and revived the idea first put forth in \citet{2007ApJ...659L..13O} that the event was a SN Ia occurring inside a dense CSM. In particular, this argument concentrated on the emission of neutral Fe lines in the late-time spectra, which those authors interpreted as being powered by radioactive decay in the nebular SN ejecta with a high Fe abundance. The striking similarities between the late-time spectra of SN~2019esa and SN~2006gy prompt us to question the hypothesis of a Type Ia-CSM  for both SNe, and is discussed in the following section.

 \begin{figure}
 \includegraphics[width=\linewidth]{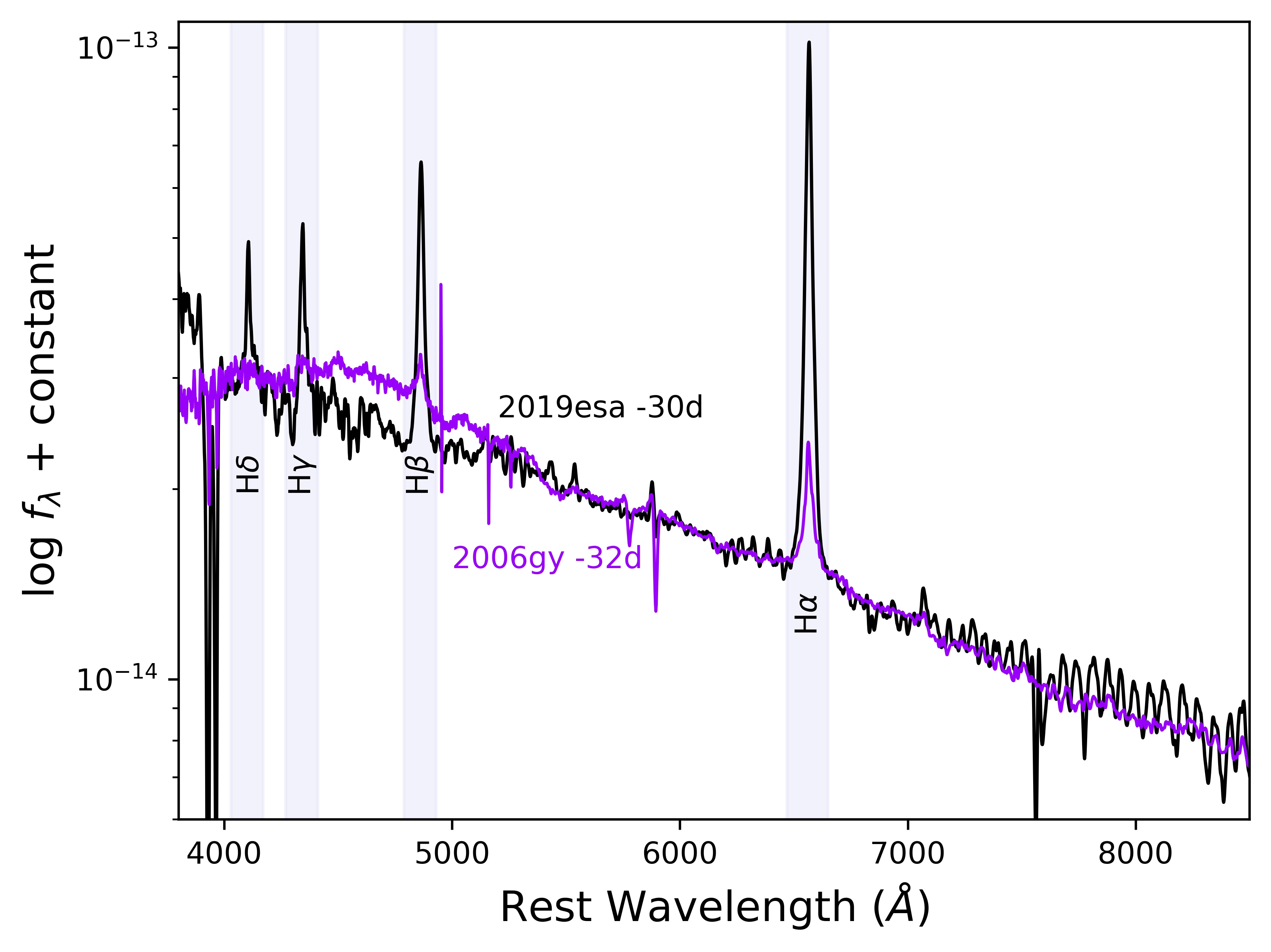}
\includegraphics[width=\linewidth]{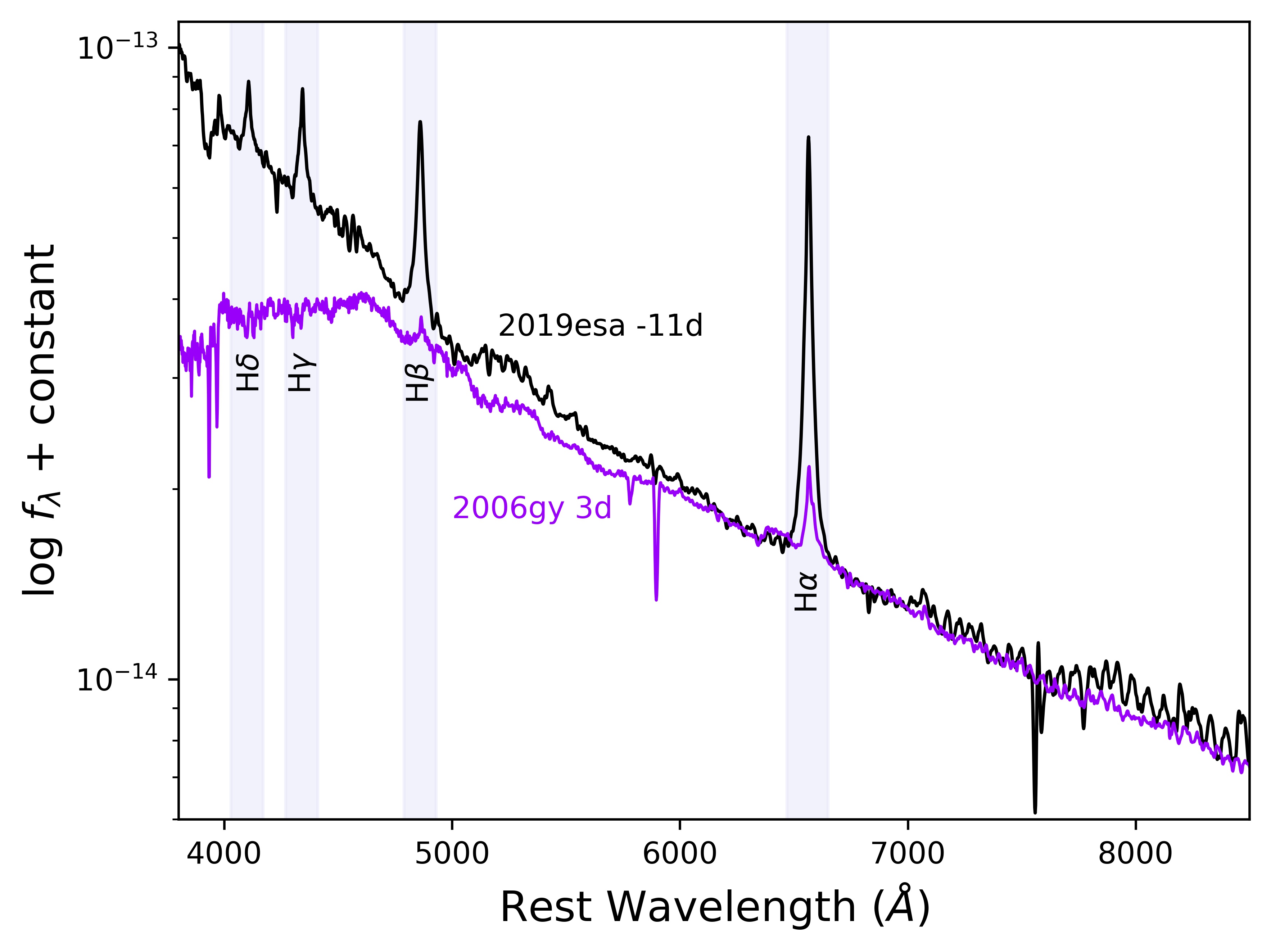}
\includegraphics[width=\linewidth]{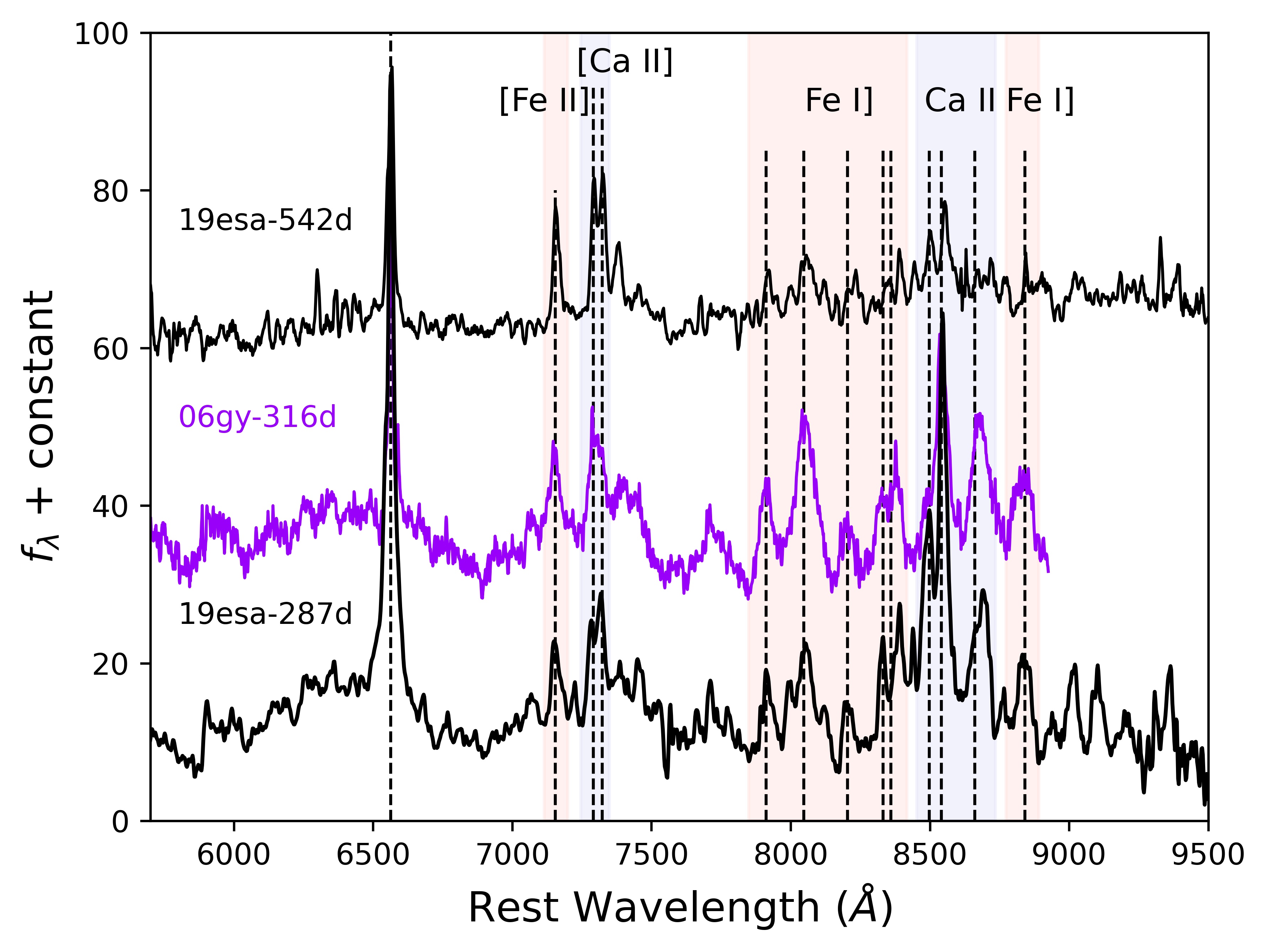}
\caption{Spectral comparison between SN~2019esa and SN~2006gy before max (top), near max (middle) and at late times (bottom). SN~2006gy spectra are from \citet{2007ApJ...666.1116S} and \citet{2009ApJ...697..747K} and have been dereddened. All epochs are with respect to the date of r/R-band maximum. }
\label{fig:Comp06gy}
\end{figure}

\subsection{Core-collapse or thermonuclear?}
SN Ia-CSM is still a poorly understood observational class, owing to the strong CSM interaction which effectively obscures the underlying SN emission. The luminosity range for Ia-CSM is between $-19.5 < M_{V} < -21.6$ \citep{2015A&A...574A..61L}, and their optical spectra show H$\alpha$ and weak H$\beta$ emission lines with generally a lack of \ion{He}{1} \citep{2013ApJS..207....3S}. More importantly Ia-CSM normally appear spectroscopically as a diluted SN Ia, then slowly begin to show narrow hydrogen emission lines which get stronger with time as the SN ejecta plows into the CSM. Only in those objects with weaker CSM interaction can the thermonuclear nature be confirmed, with PTF11kx (bottom panel, Figure \ref{fig:spectracompare}) being the most unambiguous case of a Ia-CSM \citep{2012Sci...337..942D,2013ApJ...772..125S,2017ApJ...843..102G} .

For SN~2019esa, we can rule out the appearance of Ia-like spectra at early times since only narrow hydrogen emission lines are seen atop a smooth blue continuum.  Furthermore, the H$\alpha$ luminosities for SN~2019esa (see Table \ref{tab:FWHM}) are much higher than the 1--9 $\times$ 10$^{40}$ erg s$^{-1}$ normally seen in Ia-CSM \citep{2013ApJS..207....3S}.  This may be due to the stronger than normal CSM interaction obscuring the underlying thermonuclear explosion or could point to a core collapse origin. Like Ia-CSM, SN~2019esa also lacks He emission lines, but this line is also missing in many core-collapse SNe IIn. With our more generous reddening correction ($E(B-V)=0.75$ mag), the absolute magnitude of SN~2019esa rose to M$_{r}$ = --19.4 mag, which is a typical peak luminosity of normal SNe Ia without CSM interaction - this makes a SN Ia-CSM model highly unlikely for SN~2019esa, because there are no SN Ia features in the spectrum near peak. Although we must reiterate that the extinction correction could be less, and the SN even less luminous, strengthening this argument. The CSM surrounding SN~2019esa is likely massive, therefore the energetics of a normal SN Ia combined with the kinematics of CSM-shock interaction would have produced a much brighter transient than was observed.  Additionally, the mass loss rates estimated for SN~2019esa and needed to sustain the CSM interaction for well over a year are way too high for thermonuclear progenitors.  For instance, the proposed SN Ia-CSM SN~2002ic had an estimated mass loss rate of $\sim$ 10$^{-4}$ M$_\sun$ yr$^{-1}$ \citep{2004MNRAS.354L..13K}, and SN~2008J was roughly 3 $\times$ 10$^{-3}$ M$_\sun$ yr$^{-1}$ \citep{2012A&A...545L...7T}. \citet{2013ApJS..207....3S}, though, estimate mass loss rates from their sample of possible SNe Ia-CSM to be much larger at a few 10$^{-1}$ M$_\sun$ yr$^{-1}$ based on rise times and X-ray observations. 

The late-time spectra may prove the best evidence we have of the true nature of SN~2019esa, in particular the presence of the [\ion{Ca}{2}] $\lambda\lambda$7291, 7324 doublet, which is normally not seen in Ia-CSM \citep{2013ApJS..207....3S}. Additionally, the \ion{Ca}{2} triplet is broad and blended in SN Ia-CSM (for example PTF11kx in Figure \ref{fig:spectracompare}), but in SN~2019esa the individual \ion{Ca}{2} components are easily resolved owing to the much lower line velocities. Most intriguing are the forest of Fe I and Fe II emission lines (Figure \ref{fig:Comp06gy}, bottom), which as we discuss next, have been proposed as evidence for a Ia-CSM nature in SN~2006gy. Other than these late-time Fe emission lines, there is no evidence that SN~2019esa is of a thermonuclear origin, and there is considerable reason to doubt such an association.

The argument for the Ia-CSM nature of SN~2006gy presented in \citet{2020Sci...367..415J} hinges on the existence of these neutral Fe lines that emerge in the late-time spectra, and the assertion that they must arise from Fe-rich nebular SN ejecta powered by radioactivity. From their analysis applying a nebular-phase SN Ia radiative transfer model, they suggest at least 0.3 M$_{\sun}$ of iron has been mixed with a 20 M$_{\sun}$ shell of H-rich CSM.  It is this dense CSM that quickly decelerated the ejecta and trapped gamma-rays, giving rise to the intermediate-width Fe lines at later times.  However, the ionization level and density used to derive this mass of Fe come from a nebular SN Ia model powered by radioactivity, even though the luminosity at late times in SN~2006gy is clearly decaying more slowly than for $^{56}$Co and is not powered by radioactivity \citep{2008ApJ...686..467S,2010AJ....139.2218M,2015MNRAS.454.4366F}.   It is also worth noting that strong Fe I and Fe II lines are seen in the CSM shells of LBVs like $\eta$ Car \citep{2004A&A...419..215H}, and do not require enhanced Fe abundances but rather unusual excitation conditions and high density.

In the bottom panel of Figure \ref{fig:Comp06gy} we show that these same Fe lines are seen in the late-time spectra of SN~2019esa, and in fact the spectra are almost identical to that of SN~2006gy.  The lower luminosity of SN~2019esa compared to SN~2006gy presents a problem for the Ia-CSM model.  Due to their homogeneous nature we expect that all SN Ia produce a similar mass of Fe.  Therefore if  both SN~2006gy and SN~2019esa have comparable Fe mass, then they should have the same underlying  radioactivity-powered lightcurve.   The lower CSM interaction luminosity of  SN~2019esa requires a smaller mass of CSM, which would create a much lower M$_{CSM}$/M$_{Fe}$ ratio compared to SN~2006gy. In this scenario, the Fe lines should therefore be much stronger in SN~2019esa compared to other lines in the spectrum, but as we see in Figure \ref{fig:Comp06gy} the late-time spectra clearly have the same relative line strengths of Fe compared to other lines.
As we presented above though, it is unlikely that SN~2019esa comes from a SN Ia, therefore the presence of neutral Fe lines in both SN~2019esa and SN~2006gy would suggest that a large Fe mass from a thermonuclear explosion is not needed to reproduce the late-time spectrum of SN~2006gy. This demonstrates that one can get the same features from a less luminous SN IIn. 

Kinematically, it would also be difficult to have a SN Ia power both SN~2006gy and SN~2019esa since the higher inferred CSM mass powering the luminosity of SN~2006gy would suggest a much slower expansion speed in the shocked gas in order to conserve momentum.   SN~2006gy has similar to somewhat faster H$\alpha$ core velocities (1500 km s$^{-1}$) than SN~2019esa, and even shows evidence of broad emission lines $>$ 4000 km s$^{-1}$ that are not seen in our SN~2019esa data. It is therefore likely that both SN~2006gy and SN~2019esa can be produced through the interaction between the explosion of a massive star and very dense CSM.

\section{Conclusions} \label{sec:Conc}
We have presented a comprehensive set of photometry and early-time spectroscopy of SN~2019esa. The unprecedented $TESS$ coverage combined with the high-cadence DLT40 lightcurve puts tight constraints on the explosion epoch and early photometric evolution.  From the observational data presented here we suggest that SN~2019esa is likely a SN Type IIn with a roughly 30 day rise time that was produced when a massive star exploded in a dense cocoon of $\sim$1--2 M$_{\sun}$ CSM produced by an eruption of the star some 3--4 years prior to explosion. The mass loss rate of $\sim$ 0.3 M$_\sun$ yr$^{-1}$ estimated from the bolometric luminosity provided by our extensive Las Cumbres Observatory photometry and the H$\alpha$ velocities derived from our optical spectra suggests that only a star with a high episodic mass loss rate, such as an LBV, is likely the type of progenitor responsible for SN~2019esa.

The presence of intermediate-width Fe emission lines in the late-time spectra of SN~2019esa are reminiscent of those seen in the SLSN SN~2006gy, suggesting that a similar mechanism responsible for SN~2006gy seems to be at play in SN~2019esa, but to a lesser degree. The fainter bolometric luminosity and brighter H$\alpha$ emission line luminosity of SN~2019esa compared to those seen in Ia-CSM point to a core collapse origin. Additionally the fact that the red end of the late-time spectra of SN~2019esa is almost identical to that of SN~2006gy, suggest that a thermonuclear engine is not needed to produce events such as these.  

With the discovery of more SN minutes to hours after explosion we are able to put tight constraints on the early time evolution of explosive events.  From pinpointing the explosion date, to searching for fleeting signatures such as high ionization lines or lightcurve undulations due to inhomogeneities in the surrounding CSM or the presence of a companion, this early time data are essential to understanding the physics and diversity among supernovae. This will become immensely important in the upcoming era of the Rubin Observatory when transients of all flavors will be discovered at an unprecedented rate. 

\begin{acknowledgments}

Supported by the international Gemini Observatory, a program of NSF's NOIRLab, which is managed by the Association of Universities for Research in Astronomy (AURA) under a cooperative agreement with the National Science Foundation, on behalf of the Gemini partnership of Argentina, Brazil, Canada, Chile, the Republic of Korea, and the United States of America. Time domain research by the University of Arizona team and D.J.S.\ is supported by NASA grant 80NSSC22K0167, NSF grants AST-1821987, 1813466, 1908972, \& 2108032, and by the Heising-Simons Foundation under grant \#2020-1864. Research by Y.D., N.M., and S.V.\ is supported by NSF grants AST-1813176 and AST-2008108. The Las Cumbres Observatory team is supported by NSF grants AST-1911225 and AST-1911151, and NASA Swift grant 80NSSC19K1639. K.A.B. acknowledges support from the DIRAC Institute in the Department of Astronomy at the University of Washington. The DIRAC Institute is supported through generous gifts from the Charles and Lisa Simonyi Fund for Arts and Sciences, and the Washington Research Foundation.  Based in part on data acquired at the Siding Spring Observatory 2.3\,m. We acknowledge the traditional owners of the land on which the SSO stands, the Gamilaraay people, and pay our respects to elders past and present.

\end{acknowledgments}

\facilities{TESS, ADS, NED, CTIO:PROMPT, Meckering:PROMPT, Las Cumbres Observatory (FLOYDS,Sinistro), Magellan:Clay (LDSS3),WISeREP}

\software{Astropy \citep{2013A&A...558A..33A,astropy}, FLOYDS pipeline \citep{Valenti14}, {\sc lcogtsnpipe} \citep{Valenti16}, IRAF \citep{IRAF}}

\bibliography{ms}
\bibliographystyle{aasjournal}

\end{document}

%% file: affiliation.tex
\newcommand{\UA}{\affiliation{Steward Observatory, University of Arizona, 933 North Cherry Avenue, Tucson, AZ 85721-0065, USA}}

\newcommand{\GNL}{\affiliation{Gemini Observatory/NSF's NOIRLab, 670 N. A'ohoku Place, Hilo, HI 96720, USA}}

\newcommand{\UW}{\affiliation {DIRAC Institute, Department of Astronomy, University of Washington, 3910 15th Avenue NE, Seattle, WA 98195, USA}}

\newcommand{\keck}{\affiliation{W.M. Keck Observatory, 65-1120 Mamalahoa Highway, Kamuela, HI 96743, USA}}

\newcommand{\LCO}{\affiliation{Las Cumbres Observatory, 6740 Cortona Drive, Suite 102, Goleta, CA 93117-5575, USA}}
\newcommand{\UCSB}{\affiliation{Department of Physics, University of California, Santa Barbara, CA 93106-9530, USA}}
\newcommand{\KITP}{\affiliation{Kavli Institute for Theoretical Physics, University of California, Santa Barbara, CA 93106-4030, USA}}
\newcommand{\UCD}{\affiliation{Department of Physics and Astronomy, University of California, Davis, 1 Shields Avenue, Davis, CA 95616-5270, USA}}

\newcommand{\CfA}{\affiliation{Center for Astrophysics \textbar{} Harvard \& Smithsonian, 60 Garden Street, Cambridge, MA 02138-1516, USA}}

\newcommand{\IAIFI}{\affiliation{The NSF AI Institute for Artificial Intelligence and Fundamental Interactions}}

\newcommand{\USask}{\affiliation{Department of Physics \& Engineering Physics, University of Saskatchewan, 116 Science Place, Saskatoon, SK S7N 5E2, Canada}}

%% file: authors.tex
\author[0000-0003-0123-0062]{Jennifer E. Andrews}
\GNL
\author[0000-0002-0744-0047]{Jeniveve Pearson}
\UA
\author[0000-0001-9589-3793]{M.~J. Lundquist}
\keck
\author[0000-0003-4102-380X]{David J. Sand}
\UA
\author[0000-0001-5754-4007]{Jacob E. Jencson}
\UA
\author[0000-0002-4924-444X]{K. Azalee Bostroem}
\UW
\author[0000-0002-0832-2974]{Griffin Hosseinzadeh}
\UA
\author[0000-0001-8818-0795]{S.~Valenti}
\UCD
\author[0000-0001-5510-2424]{Nathan Smith}
\UA
\author[0000-0002-1546-9763]{R.~C. Amaro}
\UA
\author[0000-0002-7937-6371]{Yize Dong \begin{CJK*}{UTF8}{gbsn}(董一泽)\end{CJK*}}
\UCD
\author[0000-0003-0549-3281]{Daryl Janzen}
\USask
\author[0000-0002-7015-3446]{Nicol\'as Meza}
\UCD
\author[0000-0003-2732-4956]{Samuel Wyatt}
\UA
\author[0000-0003-0035-6659]{Jamison Burke}
\LCO\UCSB
\author[0000-0002-1125-9187]{Daichi Hiramatsu}
\CfA\IAIFI
\author[0000-0003-4253-656X]{D.\ Andrew Howell}
\LCO\UCSB
\author[0000-0001-5807-7893]{Curtis McCully}
\LCO\UCSB
\author[0000-0002-7472-1279]{Craig Pellegrino}
\LCO\UCSB

%% file: Speclog.tex
 \begin{deluxetable*}{lcccccc}
\tablecaption{Optical Spectroscopy of SN~2019esa \label{tab:optspec}}
\tablehead{ \colhead{UT Date}    &\colhead{MJD}& \colhead{Phase}    &\colhead{Telescope+}   & \colhead{R}&  \colhead{Exposure Time}  \\ 
   \colhead{(y-m-d)}    &\colhead{} & \colhead{(days)} & \colhead{Instrument}  &\colhead{$\lambda$/$\Delta\lambda$}   & \colhead{(s)}   \  }
\startdata
2019-05-06 & 58609.38 & 1 &  FTS+FLOYDS & 400 & 2700 \\
2019-05-08 & 58611.45 & 3 &  FTS+FLOYDS & 400 & 2700 \\
2019-05-09 & 58612.39 & 4 &  FTS+FLOYDS & 400 & 2700 \\
2019-05-14 & 58617.43 & 9 &  FTS+FLOYDS & 400 & 2700 \\
2019-05-23 & 58626.37 & 18 &  FTS+FLOYDS & 400 & 2700 \\
2019-05-26 & 58629.35 & 21 &  FTS+FLOYDS & 400 & 1800 \\
2019-06-15 & 58649.34 & 41 &  FTS+FLOYDS & 400 & 1800 \\
2020-03-18 & 58926.45 & 318 &  FTS+FLOYDS & 400 & 3600 \\
2020-11-26 & 59180.26 & 572 &  Magellan+LDSS3 & 750 & 1500 \\
\hline
\enddata
 \tablecomments{Phases are reported with respect to an explosion epoch of MJD 58608.44}
 \end{deluxetable*}